\documentclass[reprint,
groupedaddress,
 amsmath,amssymb,
prl
]{revtex4-2}

\usepackage{graphicx}
\usepackage{dcolumn}
\usepackage{bm}
\usepackage{xcolor}

\usepackage{float}
\usepackage{makecell}

\begin{document}

\title{Shape transitions of sedimenting confined droplets and capsules: from oblate to bullet-like geometries}

\author{Danilo P. F. Silva$^{1,2}$}
\author{Rodrigo C. V. Coelho$^{1,2,3}$}%
\author{Ariel Dvir$^{4}$}
\author{Noa Zana$^{4}$}
\author{Margarida M. Telo da Gama$^{1,2}$}
\author{Naomi Oppenheimer$^{4}$}
\author{Nuno A. M. Araújo$^{1,2}$}%
\affiliation{%
 $^{1}$Centro de Física Teórica e Computacional, Faculdade de Ciências, Universidade de Lisboa, P-1749-016 Lisboa,}
\affiliation{%
 $^{2}$Departamento de Física, Faculdade de Ciências, Universidade de Lisboa, P-1749-016 Lisboa,
Portugal}
\affiliation{$^3$Centro Brasileiro de Pesquisas Físicas, Rua Xavier Sigaud 150, 22290-180 Rio de Janeiro, Brazil}
\affiliation{%
 $^{4}$School of Physics and the Center for Physics and Chemistry of Living Systems, Tel Aviv University, Tel Aviv 6997801, Israel\\
}%

\begin{abstract}
The transport and deformation of confined droplets and flexible capsules are central to diverse phenomena and applications, from biological flows in microcapillaries to industrial processes in porous media. \textcolor{black}{Inspired by experiments}, we perform numerical simulations to investigate their shape dynamics under varying levels of confinement and particle flexibility. A transition from an oblate to a bullet-like shape is observed at a confinement threshold, independent of flexibility\textcolor{black}{, which agrees with our analytical calculations}. A fluid-structure interaction analysis reveals two regimes: a pressure-dominated and a viscous-dominated regime. For highly flexible particles, the pressure-dominated regime prevails and the deformation is enhanced. These findings offer new insights into the transport of flexible particles in confined environments, with implications for biomedical applications, filtration technologies, and multiphase fluid mechanics.
\end{abstract}

\maketitle

The sedimentation of particulates is a key process observed in numerous natural and industrial processes, including raindrop formation, air purification, water treatment, clinical diagnostics, and wastewater management~\cite{Szakall_Mitra_Diehl_Borrmann_2010,Moragues2023,Tishkowski_Gupta_2023,Matko_Fawcett_Sharp_Stephenson_1996}. This phenomenon is governed by factors, such as particle density, size, shape, and the viscosity and density of the surrounding fluid~\cite{ghosh_stockie_2015,Vosse_Sherwin_2007,Wang_Guo_Mi_2014}. Confinement strongly affects the dynamics of these processes, defining the strength and spatial distribution of the drag forces, as demonstrated in different studies~\cite{Takaisi_1955,Pianet_Arquis_2008,Chhabra_Agarwal_Chaudhary_2003,BenRichou_Ambari_Lebey_Naciri_2005}. Previous works have explored hydrodynamics and shape transitions in vesicles and bubbles in confined flows, highlighting the effects of spatial constraints~\cite{Kahali2024, Barakat_Shaqfeh_2018,Muradoglu_Stone_2007,Freund_2014,Tiribocchi2023,arxiv.2204.10059, doi:10.1126/science.aaw8719,doi:10.1073/pnas.2313755120, Lu_Guo_Yu_Sui_2023, Jing2024, mi11020201}. In vesicles, shape transitions are dominated by bending elasticity, as in Ref.~\cite{Huang_Abkarian_Viallat_2011}. By contrast, in droplets and capsules the dominant restoring mechanism is interfacial tension or stretching, leading to a different effective elasticity~\cite{Chaikin2000, Safran1994}. This distinction is relevant in the interpretation of our results.

While the flow of flexible bodies has been extensively studied, their sedimentation remains a relatively underexplored area of research. In these systems, the flow deforms the body which in turn changes the fluid flow around it. The impact on the settling velocity of flexible particles has significant implications for several industrial problems such as slurry mixing, food production, and fluidized bed reactors~\cite{Acrivos_1979}. Moreover, sedimentation often occurs in complex environments. For example, when a person inhales particulates (dust, aerosol), they will deposit in the lungs, a highly confined network of intricate tubes. Specifically, in the alveolus region, the Reynolds number can drop to  $\approx 0.01$~\cite{Chapter3} where sedimentation becomes the primary mode of deposition. Consequently, understanding the process is essential to advancing inhaled aerosol therapeutics~\cite{Mortensen_Hickey_2014,GHANEM2025421,Mallik_Mukherjee_Panchagnula_2020}.

The primary difference between capsules and droplets lies in their bounding interfaces: capsules have elastic solid membranes with no-slip conditions, whereas droplets have liquid interfaces with continuous tangential velocity and strain rate~\cite{D3SM01648J,Misbah_2012}. Capsule flexibility is governed by solid membrane elasticity, achieved through interfacial polymerization of a liquid droplet, while droplet flexibility depends on surface tension~\cite{Champagne_Fustier_2007,Zhu_Rabault_Brandt_2015}. Studies on sedimenting droplets and capsules revealed a range of stationary shapes~\cite{Machu_Meile_Nitsche_Schaflinger_2001,Boltz_Kierfeld_2015}, including red blood cells showing distinct shape transitions~\cite{Peltomaki_Gompper_2013, Recktenwald2022, doi:10.1073/pnas.0504243102}. However, despite significant progress in understanding these dynamics, the effect of confinement on shape transitions remains poorly understood, even for simple spherical shapes.

Here, we study the dependence of particle shape on flexibility and confinement during sedimentation in a viscous fluid. Through a combination of experiments and simulations, we observe a shape transition in confined droplets from spherical to a bullet-like shape, driven by the degree of confinement. To further investigate these transitions, we employ fluid dynamics simulations, which reveal a broader range of shapes. Our results indicate that particle shape varies significantly with both confinement and flexibility. The transition from spherical to bullet-like shape is observed under high confinement in both experiments and simulations. In contrast, for low confinement, simulations predict an additional transition from spherical to oblate shape when flexibility is significantly reduced.

In the experimental setup, glycerol droplets of varying volumes ($V = 12$ – $22$ $\mu$L), mixed with a dye, are released at the top of a narrow circular channel with a diameter of $W = 5$ mm and a length of $L = 40$ cm, filled with silicone oil (viscosity $0.06 \,{\rm Pa} \cdot{\rm s} $) (see Supplementary Material (SM), Fig. S1). \textcolor{black}{The droplet volume was varied in order to tune the confinement ratio 
$k = D/W$, since the undeformed droplet diameter $D$ increases with volume.} The velocity and contour of each droplet are tracked during sedimentation. The Reynolds number ($Re$) is estimated to range between $[0.04, 0.09]$, and the confinement ratio $k$, defined as the ratio of particle to tube diameters, is in the range $[0.01, 0.7]$. Surface tension $\sigma$ is adjusted by adding surfactants in varying concentrations, yielding values of 4--20 mN/m, measured using the pendant drop technique (see Supplementary Material~\cite{ de2004special}). No visible changes in droplet shape or size were observed in the steady-state configuration. Droplets under low confinement remained spherical, whereas those under high confinement adopted a bullet-shaped contour. These shapes are consistent with results reported in the literature~\cite{Borhan_Pallinti_1995}. Results for high confinement are presented, as low confinement showed no noticeable deviation from sphericity (see Fig.~\ref{fig:shapetransitions}a, top row). {\color{black} The glycerol solution was mixed with Tween 80 (Sigma-Aldrich) as surfactant and a commercial
food dye (A.L. Spices Manufacturing \& Marketing Ltd.) for visualization. The mixture was stirred
magnetically for 5 minutes prior to injection. The dye was used in small concentrations, and no
qualitative differences in droplet behavior were observed compared to undyed solutions, consistent
with the expectation that surface tension variations were negligible at these concentrations.}

\begin{figure}
  \center
  \includegraphics[width=0.8\columnwidth]{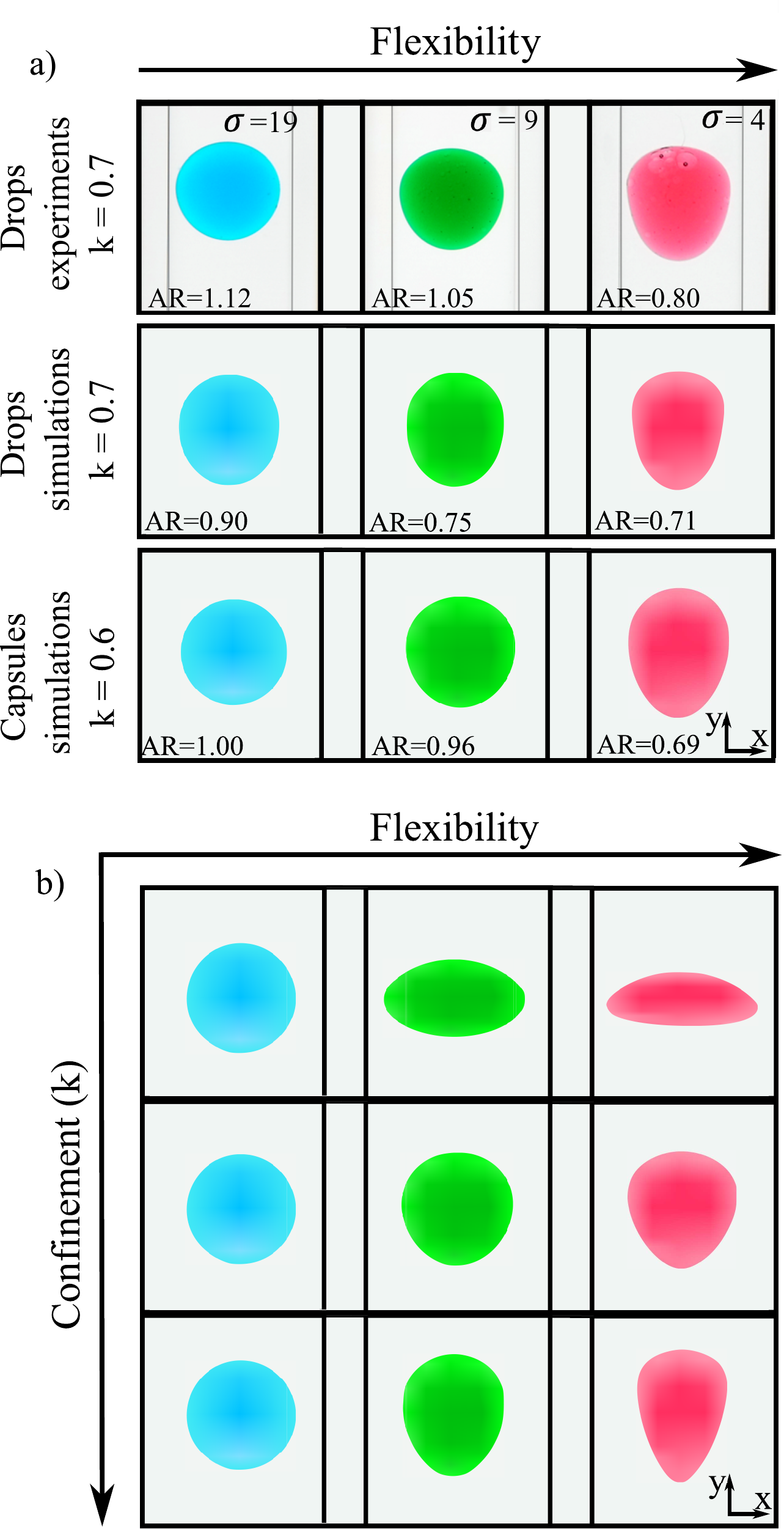}
  \caption{{(a) Shapes of sedimenting droplets (experiments and simulations) and capsules (simulations) at high confinement $k$. In the first row, we show droplets of dyed glycerol sedimenting in silicone oil. We use different volumes $12 \mu$L-- $22 \mu$L and surface tensions $\sigma$ $\{19, 9, 4\}$ mN/m (left to right). Whereas the small droplets with high surface tension remain spherical, the larger droplets with the lowest surface tension become bullet-shaped (pink colour). The grey vertical lines in the first row are the tube edges. AR stands for aspect ratio of the steady shape. (b) Shape transitions for capsules for different values of confinement parameter $k$ \textcolor{black}{(between 0.05 and 0.6)} and Bond number $\mathrm{Bo}$} \textcolor{black}{between 0.079 and 790} obtained numerically \textcolor{black}{(the parameters are available in the SM)}.
  }
  \label{fig:shapetransitions}
  \end{figure}

To achieve better control over system conditions, parameters, and measurements, we extended this study by analyzing the shape transitions using numerical simulations. Specifically, we implemented lattice Boltzmann simulations of sedimenting droplets, which yielded \textcolor{black}{the bullet-like shape} consistent with the experimental results, as shown in Fig.~\ref{fig:shapetransitions}a (middle row). More details of these simulations can be found in the SM and Refs.~\cite{Silva_Coelho_daGama_Araujo_2023,coelho2023}. However, both the droplet simulations and experiments are limited to a relatively small range of flexibilities. To overcome this limitation, we also performed simulations of flexible capsules, combining the lattice Boltzmann method with the immersed boundary method to couple the capsule dynamics with hydrodynamics~\cite{nakamura_spring-network-based_2013,nakamura_analysis_2014,wu_simulation_2013}. Simulating flexible capsules allows us to explore parameter regions corresponding to high flexibility and low confinement, which are inaccessible for droplets. Moreover, measuring the force distribution from different contributions is significantly easier in capsule simulations than droplet simulations.  For flexible capsules, we observed the same qualitative behavior under high confinement and low flexibilities, as shown in Fig.~\ref{fig:shapetransitions}a (bottom row). Therefore, for the remainder of this Letter, we focus on simulations of capsules, leveraging their technical advantages to analyze the stresses acting on the capsule and understand the mechanisms behind the shape transitions.

 The sedimentation of a flexible capsule is studied in a two-dimensional (2D) domain of
size $L \times W$, where $L$ is the length of the channel (sedimentation direction) and
$W$ is its width (confinement direction). The domain is bounded by no-slip walls at
$y=0$ and $y=W$, while periodic boundary conditions are applied in the sedimentation
direction. The solid membrane of the capsule is discretized into $40$ nodes, and a constant
body force is applied to each node, driving the sedimentation with Reynolds numbers in
the range $Re \in [0.049, 0.33]$. No-slip boundary conditions are enforced on both the
capsule boundary and the channel walls. The capsule is initially positioned at
$(x_0, y_0) = (0.5W, 0.89L)$. Since we simulate a 2D flexible particle (cylinder), the
Stokes law for 3D spheres does not apply in the low-confinement limit. Instead,
two-dimensional approximations are used to quantify the drag force, as detailed in
Ref.~\cite{ghosh_stockie_2015}. \textcolor{black}{We limited our study to capsules with equal internal
and external fluid viscosity and density; other ratios may be explored in future work.}

Under low confinement and high flexibility, the capsules undergo an additional shape transition, shifting from a circular shape at low flexibility to an oblate shape at high flexibility. Figure~\ref{fig:shapetransitions}b summarizes the observed shapes as a function of flexibility and confinement for capsules. These shapes are explored across a wide range of confinement and flexibility parameters. Next, we analyze how deformation and confinement influence the hydrodynamic stresses acting on the capsule and their impact on the resulting drag force.

\begin{figure}
\center
\includegraphics[width=1.0\linewidth]{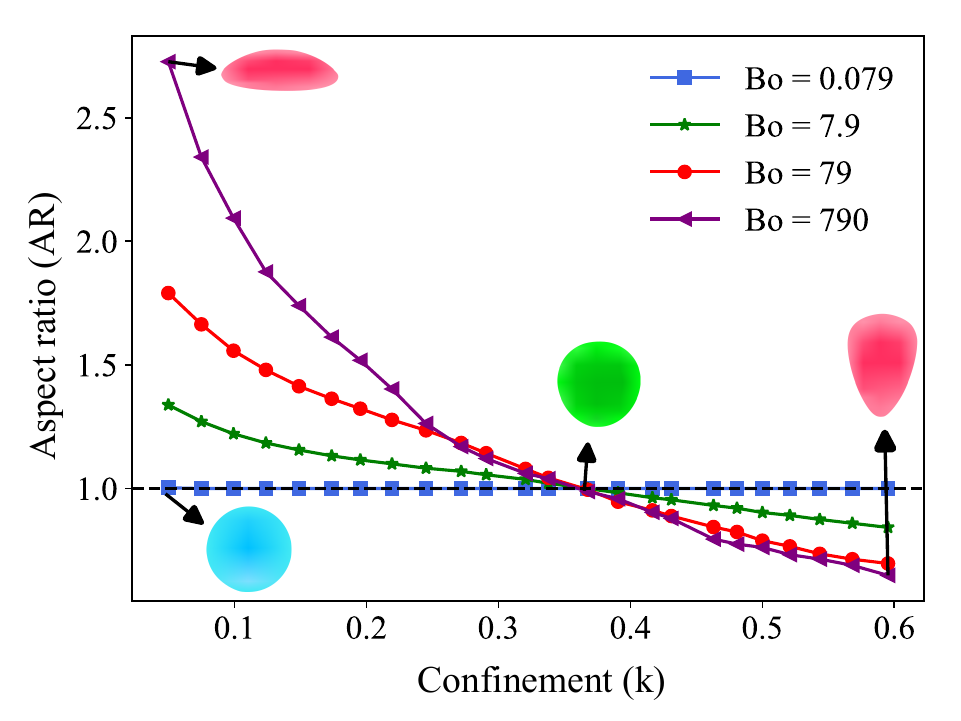}
\caption{Shape transitions for capsules differing in flexibility, determined by the Bond number $\mathrm{Bo}$. Higher $\mathrm{Bo}$ corresponds to higher flexibility. The aspect ratio AR increases with the confinement parameter as   $k \to 0$ (unconfined capsule) and decreases as $k \to 1$ (ultraconfinement). For illustration, we show different capsule shapes at different flexibilities and levels of confinement. The green capsule is on the green curve for Bo = 7.9.}
\label{fig:aspectratiovsk}
\end{figure}

The problem is characterized by two dimensionless numbers: the confinement ratio $k = \frac{D}{W}$ and
the Bond number $\mathrm{Bo}=\frac{\Delta \rho \: g \: D^2}{k_s}$. The confinement ratio $k$ represents the ratio of the \textcolor{black}{underformed} capsule diameter $D$ to the channel width $W$, while the Bond number $\mathrm{Bo}$ quantifies the relative influence of gravitational forces to elastic forces (or surface tension in the case of droplets). Here, $\Delta\rho$ is the density mismatch between the capsule and the surrounding fluid, $g$ is the gravitational acceleration, and $k_s$ is the elastic spring constant. In our capsule model, the elastic spring constant \(k_s\) represents the stretching (tension) elasticity of the membrane. For vesicles, by contrast, the dominant elasticity is bending, characterized by a bending modulus \(\kappa\). In this case, the effective stiffness is scale-dependent and combines bending and tension contributions~\cite{Chaikin2000, Safran1994}. The present study focuses exclusively on the tension-dominated regime appropriate for capsules and liquid droplets. In the simulations, the particle size $D$ and density difference $\Delta\rho$ are held constant, meaning that higher Bond number corresponds to higher particle flexibility.  

The simulations are performed at various Bond numbers and confinement levels $k$. A flexible capsule evolves into a stationary shape, and snapshots of these steady-state shapes are shown in Fig.~\ref{fig:shapetransitions}b. Figure~\ref{fig:aspectratiovsk} illustrates the dependence of the aspect ratio $\mathrm{AR}$ on $k$ for different Bond numbers. At high confinement ($k \to 1$), the capsules adopt a bullet-like shape, consistent with the behavior observed for droplets in both experiments and simulations. By contrast, at low confinement ($k \to 0$) and high flexibility, the capsules evolve to an oblate shape. The transition from an oblate-like to a bullet-like shape occurs at $k \approx 0.37$, regardless of the Bond number. For $\mathrm{Bo} = 0.079$, the capsule exhibits negligible shape changes and retains its initial circular configuration throughout the simulation. To investigate the mechanisms driving these shape transitions and the observed independence from flexibility, we analyzed the drag force and its individual contributions.

\begin{figure}
  \center
  \includegraphics[width=1.0\linewidth]{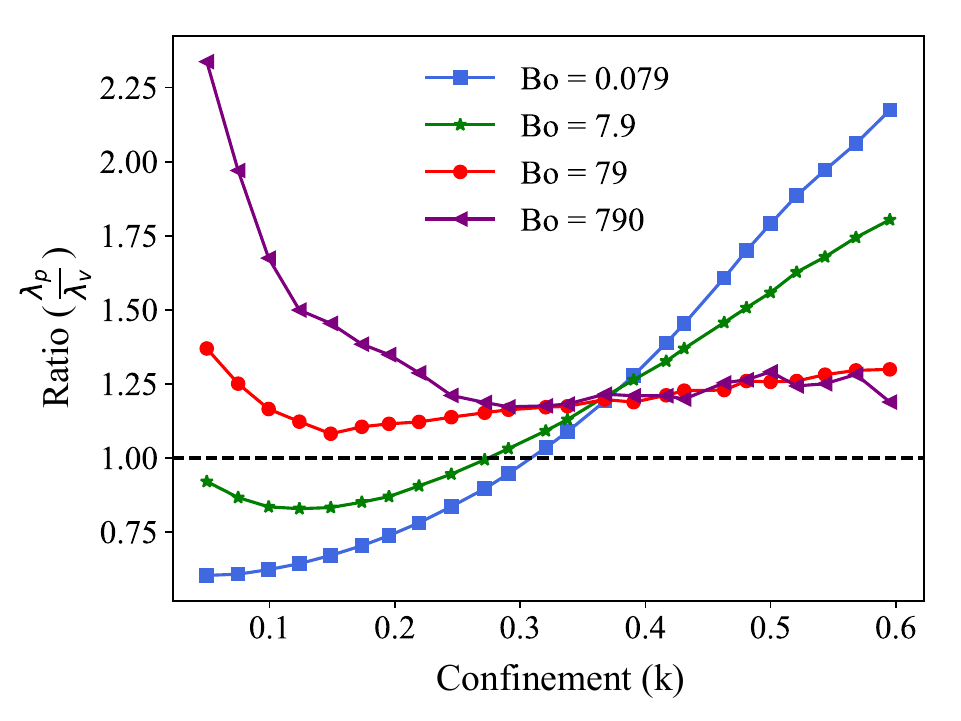}
  \caption{Ratio between the pressure and viscous contributions to the drag force. Here $ \lambda_p =F^{D}_{p} /(\mu V_t)$, $\lambda_v = F^{D}_{v} /(\mu V_t)$ and $\lambda_t = \lambda_p + \lambda_v$. The subscripts $p$, $v$ and $t$ stand for pressure, viscous and total. 
  } 
  \label{fig:lambda}
  \end{figure}

In order to quantify the influence of hydrodynamic forces on the drag force $F^D$, we analyzed the fluid stresses acting on the capsule. The motion of the capsule drives fluid flow, which in turn affects both the motion of the capsule and its shape. Changes in the shape alter the velocity gradients and pressure distribution, leading to a dynamic interplay between the capsule and the surrounding fluid. 

The fluid stress tensor is defined as
\begin{equation}
    \sigma_{i j}= - p \delta_{ij} +\mu\left(\frac{\partial u_{i}}{\partial x_{j}}+\frac{\partial u_{j}}{\partial x_{i}}\right) , 
\end{equation}
where $p$ is the pressure, and the second term represents the viscous (or deviatoric) stress tensor, which is zero for static fluids. We calculated the pressure and viscous stresses across the capsule boundary and obtained the ratio of their contributions to the drag force (re-scaled by dynamic viscosity and terminal velocity) as $\lambda_p =F^{D}_{p} /(\mu V_t)$ and $\lambda_v = F^{D}_{v} /(\mu V_t)$, where the subscripts $p$ and $v$ stands for pressure and viscosity. To quantify the transition, we calculate the ratio of these contributions $F_{ratio}^{D}=\frac{F_{p}^{D}}{F_{v}^{D}} = \frac{\lambda_p}{\lambda_v}$. In Fig.~\ref{fig:lambda}, we observe a transition from a viscous-dominated regime to a pressure-dominated regime for capsules with lower $\mathrm{Bo}$, marked by the crossing of the dashed line at $F^{D}_{\text{ratio}} = 1$. \textcolor{black}{The transition between viscosity-dominated and pressure-dominated regimes may be characterized by the ratio of pressure to viscous contributions to drag: $\lambda_p/\lambda_v < 1$ corresponds to viscosity-dominated behavior, while $\lambda_p/\lambda_v > 1$ signals pressure-dominated behavior.} It should be noted that the curves in Fig.~\ref{fig:lambda} cross at the same point as those in Fig.~\ref{fig:aspectratiovsk}, i.e., at $k=0.37$, where the shape transition occurs. At lower $\mathrm{Bo}$ (lower flexibility), the pressure contribution increases significantly more than the viscous one as $k \to 1$, resulting in flexible capsules being dominated by pressure forces. When $k \to 0$, the pressure contributions arise from the deformations of the capsule, as the wall effects are negligible in this regime. For quasi-rigid capsules under low confinement ($k \to 0$), the walls are the primary contributors to the changes in drag and its viscous and pressure components. However, for flexible capsules, shape changes significantly alter both the pressure and viscous contributions. This interplay between shape deformation and confinement leads to drag forces for capsules with higher $\mathrm{Bo}$ being dominated by pressure stresses, while at lower $\mathrm{Bo}$, viscous stresses dominate. Individual pressure and viscous contributions as a function of confinement are shown in Fig. S8 of the Supplementary Material.

{\color{black} To provide an analytical understanding of this crossover, we performed a simple lubrication-based scaling for a rigid cylinder confined between two walls (see Sec.~VI in the Supplementary Material). The characteristic lubrication pressure scales as $P_c \sim \mu U \sqrt{2a/h_0^3}$, while the viscous shear scales as $\tau_{\mathrm{visc}} \sim \mu U / h_0$, where $a = D/2$ is the cylinder radius and $h_0 = (W-D)/2$ the half-gap to the walls. Equating these contributions yields $h_0 \sim 2a$, or in terms of confinement $k = D/W$, a critical value $k_c \sim 1/3$. This estimate is in good agreement with the numerically observed crossover $k \approx 0.37$, supporting the interpretation relating the transition mechanism to the change from viscous-dominated to pressure-dominated drag regimes.
}

We measured the forces acting on each node of the capsule, calculated from the stress tensor, as shown in Fig.~\ref{fig:traction} for four combinations of Bond numbers ($\mathrm{Bo}$) and confinement parameters ($k$), where the settling velocity is directed downwards. In all cases, the forces on the top surface point outward, while those on the bottom surface point inward \textcolor{black}{(relative to the surface)}. The force distributions differ both quantitatively and qualitatively at the two levels of confinement, $k = 0.05$ and $k = 0.6$. First, we consider the case of low flexibility ($\mathrm{Bo} = 0.079$), shown in Figs. ~\ref{fig:traction}a and b. At low confinement ($k=0.05$, Fig.~\ref{fig:traction}a), the forces are nearly parallel to the direction of motion on both sides of the capsule, but their magnitudes differ: the forces on the bottom are larger, resulting in vertical compression. By contrast, at high confinement ($k=0.6$, Fig.~\ref{fig:traction}b), the forces are primarily normal to the capsule surface \textcolor{black}{, as can be seen by visual inspection (except on the sides).}. This arises from differences in the hydrodynamic regimes. At low confinement, the viscous-dominated regime prevails, where velocity gradients acting on the top and bottom are the primary contributors to the forces. At high confinement, strong pressure gradients develop near the walls and act normally across the capsule boundary, dominating the velocity gradients (see Fig.S5 in the SM). These forces can be calculated analytically in the lubrication limit for a hard cylinder sedimenting near a wall (see the SM). The cylinder creates an antisymmetric pressure (high in its front and low in its wake), acting to deform it. It can be shown that in this limit, the shear stresses are of order $(1-k)^2$ lower than the pressure.
Although this change in force distribution has minimal impact on the capsule shape at low Bond number, it has a pronounced effect at high Bond numbers. 

Now consider the case of high flexibility ($\mathrm{Bo}=790$) shown in Figs.~\ref{fig:traction}c and d. At low confinement ($k=0.05$, Fig.~\ref{fig:traction}c), as the forces act along the direction of motion, they cause the capsule to compress vertically. The bottom becomes flatter than the top due to the differences in force magnitudes. At high confinement ($k = 0.6$, Fig.~\ref{fig:traction}d),  bottom compression and top extension are observed, but the deformation differs because the forces act predominantly normal to the capsule surface. This analysis explains the shape transition shown in Fig.~\ref{fig:aspectratiovsk} and also why the threshold of confinement does not depend on the flexibility, but only on the hydrodynamic regime.

\begin{figure}
  \center
  \includegraphics[width=1.00\linewidth]{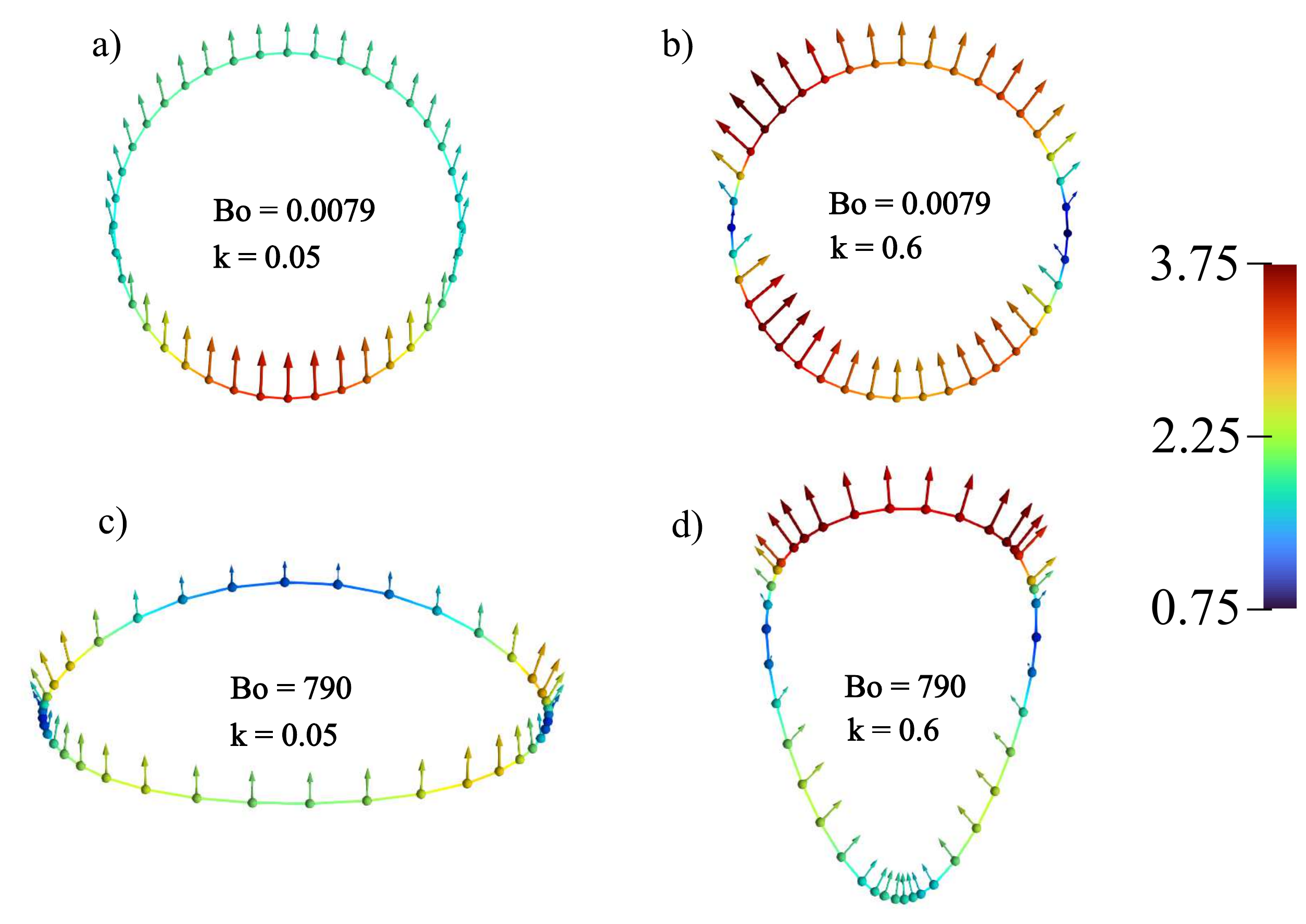} 
  \caption{Force vector and steady shape for different values of confinement and Bond number. The force vectors are scaled according to their magnitude. The results are in lattice units. The calculation of the force vectors from fluid stresses is detailed in the SM. The colorbar indicates the force magnitude, whose values are multiplied by $10^{-4}$.}
  \label{fig:traction}
\end{figure}

In conclusion, we investigated the shape transitions of sedimenting droplets and capsules in viscous fluids across varying levels of confinement and flexibility. Our findings reveal that, at high confinement, both droplets and capsules adopt a bullet-like shape, while at low confinement and high flexibility, capsules transition to an oblate shape. Using experiments and simulations, we found similar behavior for droplets and capsules, validating the observed shape transitions. However, capsules allowed us to explore a broader range of flexibilities and offered greater precision in analyzing force distributions, making them the focus of our quantitative analysis.

Two hydrodynamic regimes were identified: a viscous-dominated regime at low confinement and a pressure-dominated regime at high confinement. For highly flexible capsules, the pressure-dominated regime prevails due to significant deformations and the fluid-mediated interactions with the confining walls. Our analysis also relates the steady-state shapes of capsules to the forces acting on their surfaces, revealing that the bottom half is under compression while the top half experiences extension. These force distributions vary with confinement and flexibility, driving the observed shape transitions.

This work provides a deeper understanding of fluid-structure interactions in confined environments and highlights the role of confinement and flexibility in particle sedimentation. The insights gained have implications for various applications, particularly in the biomedical field. For instance, understanding the interplay between cell shape and sedimentation may inform new diagnostic tools, such as enhanced methods for measuring erythrocyte sedimentation rates, or contribute to the development of novel clinical tests for assessing cell health. Beyond biomedical applications, our results will inform the design of filtration systems and improve models for sedimentation in industrial processes.

Supplementary Material - For more details about the numerical method and experiments, please check the SM.

DPFS, RCVC, MMTG and NAMA acknowledge financial support from the Portuguese Foundation for Science and Technology (FCT) under the contracts: UIDB/00618/2020 (DOI:10.54499/UIDB/00618/2020), UIDP/00618/2020 (DOI:10.54499/UIDP/00618/2020), DL57/2016/CP1479/CT0057 (DOI:10.54499/DL57/2016/CP1479/CT0057), 2020.08525.BD and 2023.10412.CPCA.A2 (DOI 10.54499/2023.10412.CPCA.A2). AD, NZ and NO acknowledge financial support from NSF-BSF with the grant number 2023624.

Data availability: The data that support the findings of this study are available from the corresponding author upon reasonable request.

\newpage

\newpage
\begin{center}
    \textbf{\large Supplementary Material -- Shape Transitions of Sedimenting Confined Droplets and Capsules: From Oblate to Bullet-like Geometries}
\end{center}
\vspace{0.5cm}
\maketitle


\section{Experiments}
\begin{figure}
  \includegraphics[width=0.7\linewidth]{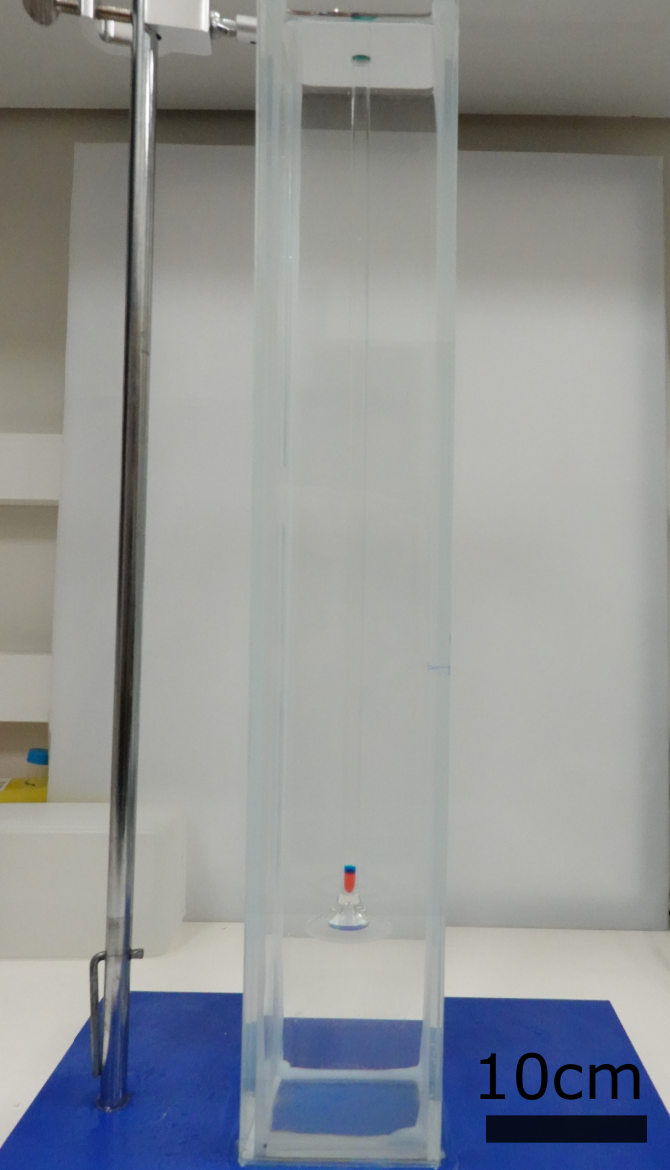}
  \caption{The main experimental setup. A water tank of dimensions $10 \times10 \times 50 {\rm cm}^3$ in which a thin capillary (diameter $5$ mm, length $40$ cm) is held.}
  \label{fig:mainSystem}
  \end{figure}

Experiments were performed using a custom built glass cylinder of an inner diameter $5$ mm and length $400$ mm. The cylinder was filled with silicone oil (viscosity $0.06 \, {\rm Pa}\cdot {\rm s}$, density 998 ${\rm kg}/{\rm m}^3$) and held in a square tank of water of dimensions $7 \times 7 \times 50 \, {\rm cm}^3$ to correct for light diffraction (see Fig.~\ref{fig:mainSystem}). Droplets were made of glycerol combined with different concentrations of detergent and a small amount of water-based food coloring. The viscosities of glycerol is 915 cP and that of glycerol + 10\% color is 246 cP.

For each experiment, the ratio between dye and glycerol was respectively 1:10. The detergent concentration was $0\, {\rm v}\%, 1 \, {\rm v}\%$ and $4 \, {\rm v}\%$. For each of these concentrations, two droplet volumes were used, $\sim 20$ and $\sim 30$ $\mu{\rm L}$. The volume of the drops was estimated using Eppendorf pipettes. Videos were processed using ImageJ. Data analysis was performed in Python using {\it openCV} library to detect the contour and volume of each drop. 

\begin{figure}
  \includegraphics[width=1.0\linewidth]{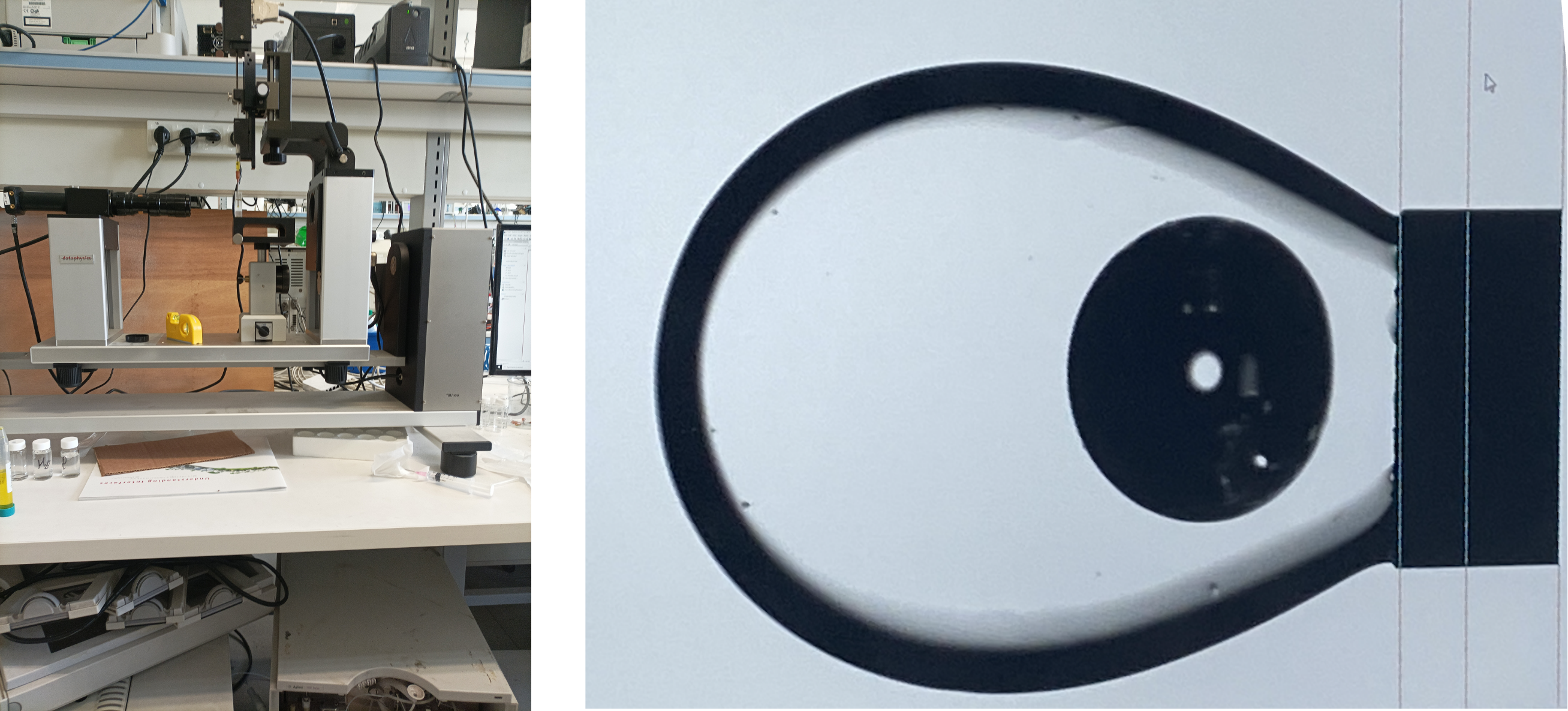}
  \caption{The optical tensiometer (left) and typical images of a suspended drop of glycerol+dye+surfactant immersed in silicone oil (right). The profile was fitted to the Young-Laplace equation to extract the surface tension. Note that the bubble in the image disappears quickly and we have waited enough to take the measurements.}
  \label{fig:tensionSystem}
  \end{figure}

Surface tension measurements were carried out using a commercial optical tensiometer (DATAPHYSICS OCA15EC, see Fig.~\ref{fig:tensionSystem}) by hanging drops of glycerol+dye+detergent in a silicone oil bath inside a glass cuvette and fitting the profile of the pendant drop to the Young-Laplace model \cite{de2004special}.

\textcolor{black}{Droplets were produced using Eppendorf Research Plus pipettes, where the target volume was set according to the values reported in the manuscript. Owing to the high viscosity of the glycerol solution, small variations in droplet volume occurred between runs. To ensure accuracy, we tracked the droplet contour with a Python routine and, assuming isotropy, estimated its volume from the measured radius. The injection process was carried out manually in a controlled manner: the pipette tip was positioned inside the oil-filled tube and the droplet was expelled as completely as possible, which guaranteed uniform release conditions. The injection speed did not affect the resulting droplet size or shape. Prior to injection, the glycerol was mixed with dye (A.L. Spices Manufacturing \& Marketing Ltd.) and Tween 80 (Sigma-Aldrich) using a magnetic stirrer for 5 minutes to obtain a homogeneous solution. The region of interest (ROI) of the camera was selected based on preliminary tests, in which we confirmed that droplets reached a steady configuration after sedimenting a few centimeters. The camera was therefore focused on this downstream section of the tube, ensuring that only droplets with stable shapes were analyzed.}

{\color{black} In the droplet experiments reported here, only nearly spherical and bullet-like steady shapes
were observed, depending on the confinement. Oblate configurations, which arise in capsules
with finite elastic modulus, are not accessible experimentally with our system because the
required Young’s modulus is too low for available materials. Such shapes are therefore only
addressed in the simulations.

We performed tests at several confinement ratios. For smaller $k$, droplets remained essentially
spherical and did not exhibit qualitatively new behavior. For stronger confinement, droplets
consistently developed the bullet-like shape reported in the main text. We therefore focus on
the intermediate confinement range, where shape transitions are most relevant.}

{\color{black}  
Throughout the manuscript and SM, the confinement ratio is defined as 
\[
k = \frac{D}{H},
\]
where \(D\) is the undeformed diameter of the particle (droplet or capsule at release) and 
\(H\) is the inner diameter of the tube. Using the undeformed size provides a consistent control 
parameter that is independent of the steady-state deformation, which varies with confinement and 
elasticity. The final deformed shape can occupy a different fraction of the cross-section, but 
for clarity and reproducibility we always report values of \(k\) based on the undeformed diameter.
}

%

\section{Simulations}

The dynamics of the fluid are governed by the NS equations given by
\begin{equation}
\begin{aligned}
&\rho\left(\frac{\partial \bm{u}}{\partial t}+(\bm{u} \cdot \nabla) \bm{u}\right)=-\nabla p+\mu \nabla^{2} \bm{u}+\bm{f_{ib}}, \\
&\nabla \cdot \bm{u}=0
\end{aligned}
\end{equation}
where $\rho$ is the fluid density, $\bm{u}$ is the fluid velocity, $p$ is fluid pressure, $\mu$ the kinematic viscosity and $\bm{f_{ib}}$ encompasses the elastic, bending, area conservation and gravity forces generated by the immersed structure on to the fluid. Instead of solving the NS equations we use the lattice Boltzmann method. We use a model for the membrane and to couple the membrane dynamics with the fluid we implemented the immersed boundary method. The assumption is made that the fluid is both Newtonian and incompressible, and that the immersed structures are also incompressible, massless, have neutral buoyancy, and occupy no space. It is also assumed that the viscous properties of the structures are the same as those of the surrounding fluid. Therefore, the system can be described by a single velocity field, which follows the incompressible Navier-Stokes equations, making the governing equations simpler. In the Immersed Boundary Method (IBM) formulation \cite{peskin_2002}, the fluid is described using an Eulerian approach, while the structures are described using a Lagrangian approach. 

\subsection{Lattice Boltzmann method}

We start with the discretisation of Boltzmann equation with the Bhatnagar–Gross–Krook (BGK) collision operator over space and time. Let $\delta x$ and $\delta t$ represent the physical distance between two adjacent lattice nodes and the time step. The discretized BGK scheme with force term is described by
\begin{multline}
\underbrace{f_{\alpha}\left(\boldsymbol{x}_{i}+\boldsymbol{\xi}_{\alpha} \delta t, t+\delta t\right)}_{\text {streaming }}=\\\underbrace{f_{\alpha}\left(\boldsymbol{x}_{i}, t\right)-\frac{\delta t}{\tau}\left[f_{\alpha}\left(\boldsymbol{x}_{i}, t\right)-f_{\alpha}^{eq}\left(\boldsymbol{x}_{i}, t\right)\right]+\mathcal{F}_{\alpha}}_{\text {collision }},
\label{eq:discrete_LBE}
\end{multline} where $\alpha$ is the index of discretized velocities, $\bm{\xi}_{\alpha}$ is the velocity vector, $i$ is the index of spatial lattice sites and $\mathcal{F}_{\alpha}$ is the force term. $f_{\alpha}^{eq}\left(\bm{x}_{i}, t\right)$ is the population distribution at equilibrium at position $\bm{x}_{i}$ and time $t$. It is not convenient to express the Lattice Boltzmann Equation (LBE) in physical units. We introduce lattice units so that $\delta x=1$, $\delta t=1$ and reference density $\rho_{ref}=1$. The equilibrium distribution $f_{\alpha}^{eq}$ is related to the local macroscale fluid velocity $\bm{u}$ and the speed of sound $c_{s}$ as
\begin{equation}
f_{\alpha}^{eq}\left(\bm{x}_{i}, t\right)=w_{\alpha} \rho\left(1+\frac{\bm{\xi}_{\alpha} \cdot \bm{u}}{c_{s}^{2}}+\frac{\left(\bm{\xi}_{\alpha} \cdot \bm{u}\right)^{2}}{2 c_{s}^{4}}-\frac{\bm{u}^{2}}{2 c_{s}^{2}}\right).
\label{eqn:eq_distribution}
\end{equation}
The fluid viscosity $\nu$ is related to the single relaxation $\tau$
\begin{equation}
\nu=c_{s}^{2}\left(\tau-\frac{1}{2}\right).
\label{eqn:viscosity_ls}
\end{equation}
We use the D2Q9 scheme for which we have $c_s^2 = 1/3$ and weights for the nine velocity vectors are $ \omega_{\alpha=0}= 4/9$, $\omega_{\alpha=1-4}=1/9$ and $\omega_{\alpha=5-8}=1/36$. The force term $F_{\alpha}$ is introduced to model external force fields, such as gravity or forces exerted by structure deformation. It can be expressed in terms of the external body force density $\rho \bm{g}$ and fluid macroscale velocity $\bm{u}$
\begin{equation}
F_{\alpha}=\left(1-\frac{1}{2\tau}\right) w_{\alpha}\left(\frac{\bm{\xi}_{\alpha}-\bm{u}}{c_{s}^{2}}+\frac{\bm{\xi}_{\alpha} \cdot \bm{u}}{c_{s}^{4}} \bm{\xi}_{\alpha}\right) \cdot \rho \bm{g}
\end{equation}
and the discretized versions of density and velocity  are
\begin{equation}
\begin{aligned}
\rho(\bm{x}, t) &=\sum_{\alpha} f_{\alpha}(\bm{x}, t), \\
\rho \bm{u}(\bm{x}, t) &=\sum_{\alpha} \boldsymbol{\xi}_{\alpha} f_{\alpha}(\bm{x}, t) + \rho\frac{\boldsymbol{g}}{2}.
\end{aligned}
\end{equation}
In the following subsections \ref{sub-constitutivemodel} and \ref{sub-immersedboundary}, we briefly describe the model used to describe the physics of a capsule and the immersed boundary method used to couple the capsule and the fluid as in Ref.  \cite{tanCharacterizationNanoparticleDispersion2016, peskin_2002}.

\subsection{Coarse Grained Model For Deformable Capsule}
\label{sub-constitutivemodel}
The energy of the capsule \cite{tanCharacterizationNanoparticleDispersion2016} is given by
\begin{equation}
\label{eq-potentialEnergy}
W\left(\bm{X}_{i}\right)=W_{\text {stretch }}+W_{\text {bend }}+W_{\text {area }},
\end{equation}
The nodal force derived from Eq. \ref{eq-potentialEnergy} and is given by
\begin{equation}
\bm{F}_{i}=-\frac{\partial W\left(\bm{x}_{i}\right)}{\partial \bm{x}_{i}}.
\end{equation}

We use a harmonic potential to model the stretching energy which is given by
\begin{equation}
W_{\text {stretch }}=\frac{1}{2} k_{s} \sum_{j=1, \ldots, N_{s}}\left(l_{j}-l_{j 0}\right)^{2},
\end{equation}
where $k_{s}$ is the stretching constant, $l_{j}$ is the length of the $j$ spring, and $l_{j 0}$ is the equilibrium spring length. The total stretching energy is summed over all the line segments. We use an exponential relationship given by $k_{s}=k_{s 0} e^{2(\lambda-1)}$, where $\lambda$ is the Bond stretch ratio defined as $\lambda=l / l_{0}$ \cite{nakamura_analysis_2014, tanCharacterizationNanoparticleDispersion2016}.

The bending energy is given by
\begin{equation}
W_{\text {bend }}=\frac{1}{2} k_{b} \sum_{j \in 1 \ldots N_{b}}\left(\theta_{j}-\theta_{j 0}\right)^{2},
\end{equation}
where $k_{b}$ is the bending constant, $\theta_{j}$ is the angle formed by the two outward surface normals of two adjacent line meshes that share the same node $j$. $\theta_{0}$ is the corresponding equilibrium angle. 

The area and mass of a 2D capsule need to be conserved. To achieve this we use the following area potential
\begin{equation}
\begin{gathered}
W_{\text {area }}=\frac{k_{g}\left(A-A_{0}\right)^{2}}{2 A_{0}},
\end{gathered}
\end{equation}
where $k_{g}$ is the global area conservation potential constants and $A, A_{0}$ are the area and equilibrium area enclosed by the solid membrane, respectively.

\subsection{Immersed boundary method}
\label{sub-immersedboundary}
The immersed structure (in this case is a capsule) is understood to be a parametric surface $X(s, t)$, where $s$ is the coordinate for the structure in a Lagrangian description and $t$ is the time \cite{tanCharacterizationNanoparticleDispersion2016, peskin_2002}. The fluid domain is described by Eulerian coordinates $\bm{x}$. The force $\bm{f}(\bm{x}, t)$ exerted by the structure on the fluid is interpolated as a source term in the momentum equation using
\begin{equation}
\bm{f}(\bm{x}, t)=\int \bm{F}(s, t) \delta(\bm{x}-\bm{X}(s, t)) d s,
\label{eqn:force_distribution}
\end{equation}
where $\bm{F}(s, t)$ is the force density for the structure. Typically, it is derived from energy density functions. $\delta(\bm{x})$ is the three-dimensional delta functions $\delta\left(x_{1}\right) \delta\left(x_{2}\right) \delta\left(x_{3}\right)$ where $x_{1}, x_{2}, x_{3}$ are the Cartesian components of the position vector $\bm{x}$. Similarly, the structure moving velocity is updated based on the local fluid velocity through interpolation using
\begin{equation}
\bm{u}(\bm{X}(s, t), t)=\int \bm{u}(\bm{x}, t) \delta(\bm{x}-\bm{X}(s, t)) d \bm{x},
\label{eqn:velocity_interpolation}
\end{equation}
where $\bm{u}(\bm{X}(s, t), t)$ is the structure moving velocity, $\bm{u}(\bm{x}, t)$ is the fluid velocity over the domain $\bm{x}$ at time $t$. $\delta(\bm{x})$ is the same function as used in Eqs (\ref{eqn:force_distribution}) and (\ref{eqn:velocity_interpolation}) essentially the velocity continuity conditions on the fluid structure interface. It is given by
\begin{equation}
\delta(x)= \begin{cases}0, \quad|x| \geq 2 \\ \frac{1}{8}\left(5-2|x|-\sqrt{-7+12|x|-4 x^{2}}\right), & 1 \leq|x| \leq 2 \\ \frac{1}{8}\left(3-2|x|+\sqrt{1+4|x|-4 x^{2}}\right), & 0 \leq|x| \leq 1\end{cases}.
\end{equation}

\textcolor{black}{Alternative numerical approaches such as the finite element method (FEM) can provide a rigorous continuum description of deformable capsules based on solid mechanics. However, FEM is typically more demanding to implement in complex fluid–structure interactions and can suffer from numerical instabilities. In contrast, the IBM models the capsule as a discrete network of nodes connected by springs, leading to a simpler mathematical formulation in which local quantities (e.g., forces) are straightforward to prescribe. This makes IBM computationally efficient and particularly well-suited to capture large deformations, where bending stabilization and the avoidance of buckling are less problematic than in continuum approaches. A limitation of IBM in the present context is the smoothing of stresses across the capsule boundary, since the thin capsule boundary is represented by linear interpolation over a few grid points, rather than as a sharp discontinuity as in boundary element methods~\cite{D3SM01648J}. Finally, three aspects of mass conservation are relevant: (i) the capsule boundary mass is negligible, (ii) the internal fluid mass is conserved by IBM with three- or four-point interpolation kernels, and (iii) an effective buoyant mass arises as a body force density difference between capsule boundary and fluid, which drives sedimentation.}

\textcolor{black}{The capsule deformation is governed by stretching and bending energies, modeled respectively through harmonic and angular potentials. The capsule membrane is discretized into Lagrangian nodes, and the interaction with the surrounding fluid is resolved using the IBM. In this approach, nodal forces are spread onto the Eulerian fluid grid, while local velocities are interpolated back to the Lagrangian points, ensuring reciprocal coupling between capsule and flow. This partitioned scheme allows the capsule deformation to feed back into the hydrodynamics. The temporal evolution of the capsule was integrated using a second-order Runge–Kutta method based on the midpoint rule.}

\subsection{Computation of fluid stresses}
\begin{figure}
  \includegraphics[width=1.0\linewidth]{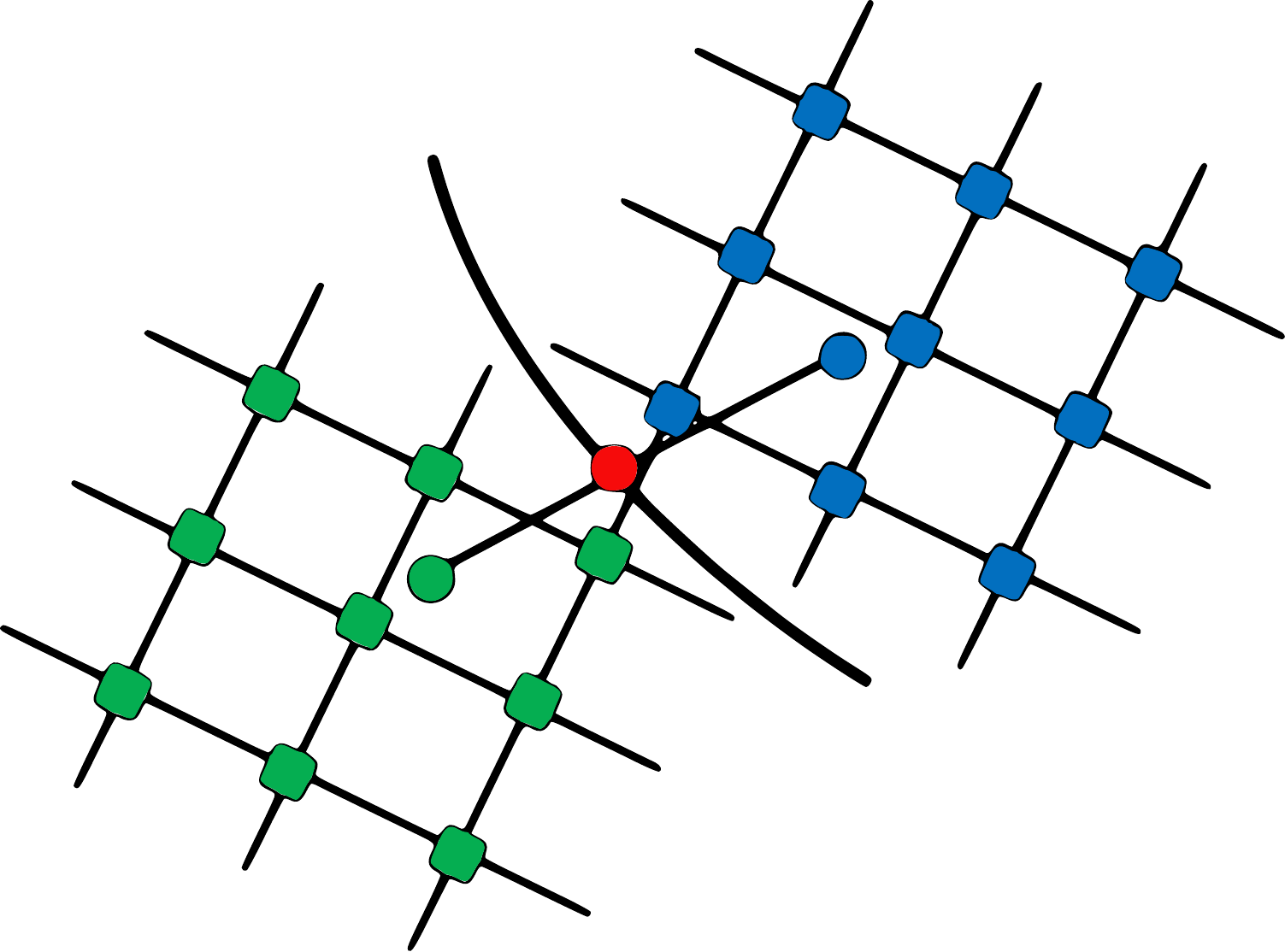}
  \caption{Calculation of the fluid stresses. There is fluid inside (blue squares) and outside (green squares) the capsule. Therefore there are pressure and viscous stresses acting across this boundary. We use the centroids (blue and green) to calculate the stress tensor inside and outside the structure.}
  \label{fig:fluidStressescalculation}
  \end{figure}

 To understand the effect of hydrodynamic forces we need to analyse the fluid stresses imparted on the capsule. The motion of the capsule generates fluid flow which in turns can affect the motion and shape of the capsule. This is because the fluid imparts fluid stresses on the capsule. The fluid stress tensor is given by
  \begin{equation}
    \sigma_{i j}= - p \delta_{ij} +\mu\left(\frac{\partial u_{i}}{\partial x_{j}}+\frac{\partial u_{j}}{\partial x_{i}}\right)
    \label{eqn:FluidStressTensor}
  \end{equation}
  where $p$ is pressure and the second term is the viscous (or deviatoric) stress tensor and only arises when there is fluid motion. Eq.(\ref{eqn:FluidStressTensor}) can also be written as $\sigma_{i j}= - p \delta_{ij} +\tau_{ij}$ where $\tau_{ij}$ is the viscous stress tensor. In matrix form
  \begin{equation}
    \left[\begin{array}{lll}
    \sigma_{xx} & \sigma_{xy}  \\
    \sigma_{yx} & \sigma_{yy} \\
    
    \end{array}\right]=\left[\begin{array}{rrr}
    -p & 0  \\
    0 & -p  \\
    \end{array}\right]+\left[\begin{array}{lll}
    \tau_{xx} & \tau_{xy}  \\
    \tau_{yx} & \tau_{yy}  \\
    \end{array}\right].
    \end{equation}
  For the viscous stress tensor, the diagonal components $\tau_{ii}$ are the viscous normal stresses and the off diagonal are the viscous shear stresses. Notice that $\tau_{xy} = \tau_{yx}$. Calculation of the velocity gradient can be done using finite differences however there is a straightforward way to calculate the stress tensor directly in LBM. The stress tensor is
  \begin{equation}
    \begin{aligned}
      \sigma_{i j} &= - p \delta_{ij} +\mu\left(\frac{\partial u_{i}}{\partial x_{j}}+\frac{\partial u_{j}}{\partial x_{i}}\right) \\
    &=-\rho c_{s}^{2} \delta_{ij}-\left(1-\frac{\Delta t}{2 \tau}\right) \sum_{\alpha}\left(f_{\alpha}-f_{\alpha}^{\mathrm{eq}}\right)\xi_{\alpha i}\xi_{\alpha j}
    \end{aligned}
    \end{equation}
  where the first term is the pressure tensor and the second the viscous tensor. Using this stress tensor we are able to calculate the fluid forces acting on the solid membrane. This approach is commonly referred to as stress integration approach as we integrate stresses to get forces. Here we focus only on the no slip boundary condition between fluid and structure. From no slip, we basically impose velocity and traction (force per unit length) continuity across the interface  
\begin{equation}
\begin{aligned}
u_{f} &=u_{s} \quad \text { on } \Omega \\
\sigma_{i j}^{f} n_{j} &=\sigma_{i j}^{s} n_{j} \quad \text { on } \Omega
\end{aligned}
\end{equation}
where $u_{f}$ and $u_{s}$ are the fluid and solid velocity. $\Omega$ is the interface between the solid and the fluid and superscripts $f, s$ represent the fluid and solid, respectively. $n_{j}$ is the surface normal. The local traction vector is then $\bm{T}=\bm{\sigma}\cdot\bm{n}$ where $\bm{n}$ is the local surface normal and its vector components are
  \begin{equation}
    \begin{gathered}
    T_{x} = \sigma_{x x} n_{x}+\sigma_{x y} n_{y} \\
    T_{y} = \sigma_{y x} n_{x}+\sigma_{y y} n_{y}.
    \end{gathered}
    \end{equation}
  This force acts normal $\bm{T_n}$ and parallel (shear) $\bm{T_s}$ to the local surface as
  \begin{equation}
    \bm{T_n} = \left( \bm{T} \cdot \bm{n}  \right) \bm{n}, \quad \bm{T_s} = \bm{T} - \bm{T_n}.
  \end{equation}
  To obtain the total force $\mathfrak{F}$ acting on the capsule we integrate the traction vector along the surface
  \begin{equation}
    \mathfrak{F}= \int_{\Omega} \bm{T} \mathrm{~d} A= \int_{\Omega} \sigma\cdot\bm{n} \mathrm{~d} S,
  \end{equation}
  where $\mathrm{~d} S$ is the elemental length (or area in 3D). Notice the $y$ component of $\mathfrak{F}$ is the drag $ F^D $ while the $x$ component is the lift $F^L$. The drag is a combination of pressure and viscous stresses acting on the capsule surface. More specifically we have pressure stresses, viscous shear stresses and viscous normal stresses. Both pressure stresses affects the shape of the capsule and act normal to the surface. Here we refer to a re-scaled drag defined as $\lambda_t = \lambda_p + \lambda_v$ where $ \lambda_p =F^{D}_{p} /(\mu V_t)$, $\lambda_v = F^{D}_{v} /(\mu V_t)$. The subscripts $t$, $p$ and $v$ stand for total, pressure and viscous. $\mu$ and $V_t$ are the dynamic viscosity and terminal velocity, respectively. We calculate the stresses acting on a boundary element as shown in Fig. \ref{fig:fluidStressescalculation}. We point out that the viscous shear stress is the component that acts tangential to the surface of a cubic fluid element in the cartesian system of coordinates while the viscous normal stress acts normal to the surface of the fluid element. This does not necessarily mean they will act tangentially and normally to the capsule surface. This would only happen if the normals of the local solid membrane and the fluid element surface match and point in the same direction. The membrane surface shear and normal is calculated using Eq. (20). Next, we validate our method for both rigid and flexible capsules.

\section{Validation of the capsule model}
\subsection{Rigid particle sedimentation}
We simulated the sedimentation of a 2D particle in a confined medium. We compared our results to analytical predictions. Following Ref. \cite{ghosh_stockie_2015}, the analytical wall-corrected terminal velocity $V_t$ of a 2D particle at $\mathrm{Re}=\frac{V_t D}{\nu}<1$ can be calculated from
\begin{equation}
V_{analytical}=\frac{\pi g D^{2}\left(\rho_{s}-\rho_{f}\right)}{4 \mu \lambda},
\label{eqn:v_c}
\end{equation}
where $D$ is the diameter of particle, $g$ is gravity, $\rho$ is the density where $s$ and $f$ stand for solid and fluid, respectively, $\mu$ is the dynamic viscosity and $\lambda$ is the wall correction factor from Eq.~\eqref{correction-eq} in the main manuscript. 

\begin{figure}
  \includegraphics[width=1.0\linewidth]{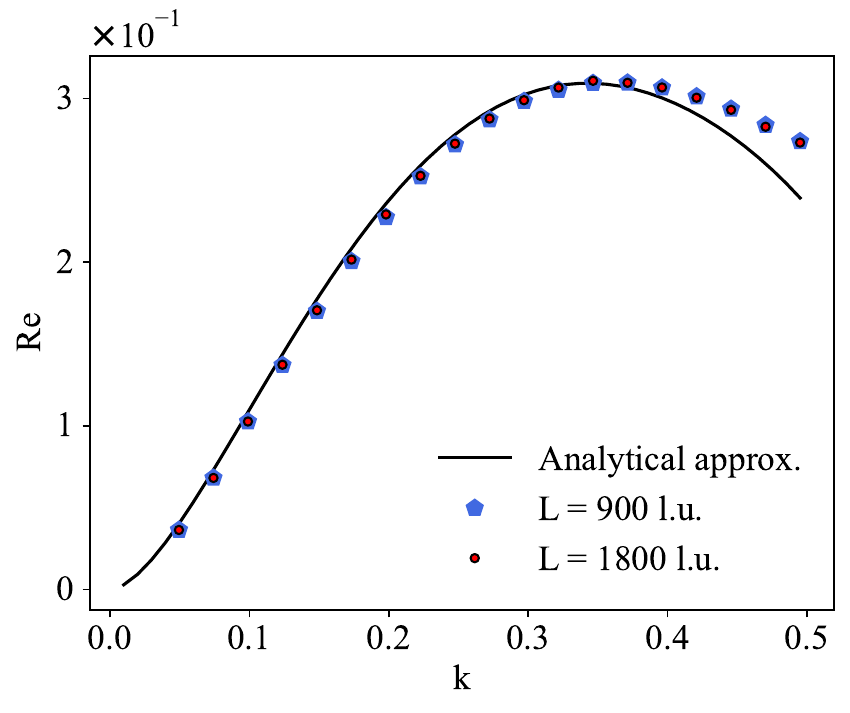}
  \caption{Quasi rigid particle validation. The simulation consists of adding a particle with a stiff membrane with $\mathrm{Bo}= 0.079$ to a rectangular channel with solid walls in both sides and let it fall driven by a constant force $g$ acting in all membrane nodes. Fluid parameters  $\rho_f = 1$ and $\tau = 1$. To check for any finite size effects the length of the channel was doubled. Convergence still occurred for $L=1800$. We fixed the width of the channel $W=100$ and the particle diameter was varied to change $k=D/H$.}
  \label{fig:rigidparticle_validation}
  \end{figure}

Confinement in a bounded fluid domain significantly impacts the drag force experienced by a sedimenting particle. This drag force reduces the settling velocity, which is captured by the wall correction factor, defined as $ \lambda(k,\mathrm{Re})=\frac{F^{D}(k)}{\mu V_t}$ where $F^D = F_y$ is the drag force per unit length and $V_t$ is the settling velocity. For low Reynolds numbers $(Re<1)$, $\lambda$ depends only on the confinement ratio $k$. In Ref.~\cite{faxen_1946}, an eighth-order wall-correction factor for a rigid particle was derived from an approximate solution of the Stokes equation, given by,
\begin{eqnarray}
    &\lambda(k) = -4 \pi/[0.9157+\ln (k)-1.724 k^{2} \nonumber \\ &+1.730 k^{4}-2.406 k^{6}+4.591 k^{8}] .
    \label{correction-eq}
\end{eqnarray}
In Fig. \ref{fig:rigidparticle_validation} we show that our results are in line with the theoretical predictions up to $k \approx 0.4$. Equation (\ref{eqn:v_c}) can correctly predict the terminal velocity up to $k = 0.5$ \cite{ghosh_stockie_2015}. We highlight that using an appropriate mesh for the particle surface is relevant to get accurate results. In our simulations, we have 40 particle surface nodes at an initial distance of 0.98 l.u. For the purposes of validation, ours simulations have $\mathrm{Re} \in [0.036,0.31]$.

\section{Further details on the numerical results}
For this supplementary material (SM) we performed simulations involving sedimentation of capsules and droplets and experiments with sedimenting droplets at different confinement levels. Our objective is to understand the shape transitions that we observe in the experiments. We performed droplet simulations however it is non trivial to measure the resultant forces and stresses at the interface of a droplet. We decided to implement a capsule model connected by nodes and springs. We found similar shape transitions and it allowed us to compute fluid stresses acting on the boundary more accurately. Nonetheless, we present our results for both droplets and capsules. The problem is characterised by two dimensionless numbers: the confinement parameter defined as $k = \frac{D}{W}$ and the Bond number $\mathrm{Bo}=\frac{\Delta \rho g D^2}{k_s}$ which measures the importance of the gravitational forces relative to the elastic forces (surface tension in the case of droplets). The behaviour of the capsule is closely related to the stresses exerted on its surface. Their evaluation provides a deeper understanding of the deformation and resultant forces acting on the capsule. From Eq. (16), (19) and (21) we see that drag will be affected by parameters $p, \tau_{yy}, \tau_{yx}$, while lift is affected by $p, \tau_{xx}, \tau_{xy}$. First let us understand the fluid motion, generated pressure and viscous stresses. We will consider the effect of confinement first and then the effect of flexibility. After, we consider the traction vectors to understand the causes of the shape change.

In our simulations, the capsules has 40 nodes at a distance of 0.98 l.u from each other with parameters $k_g = 5.5 \,10^{-4}$ and $k_b= 2.4 \,10^{-1}$ in l.u. In Fig. 1b of the main manuscript, we use 3 values of confinement $k$ which are 0.05, 0.37 and 0.59 and 3 values for the elastic constant $k_s$ which are $1.6\,10^{-5}$ (pink capsule), $1.6\,10^{-3}$ (green capsule) and $1.6 \, 10^{-1}$ (blue capsule).

\subsection{Effect of confinement}

\begin{figure}
  \includegraphics[width=1.0\linewidth]{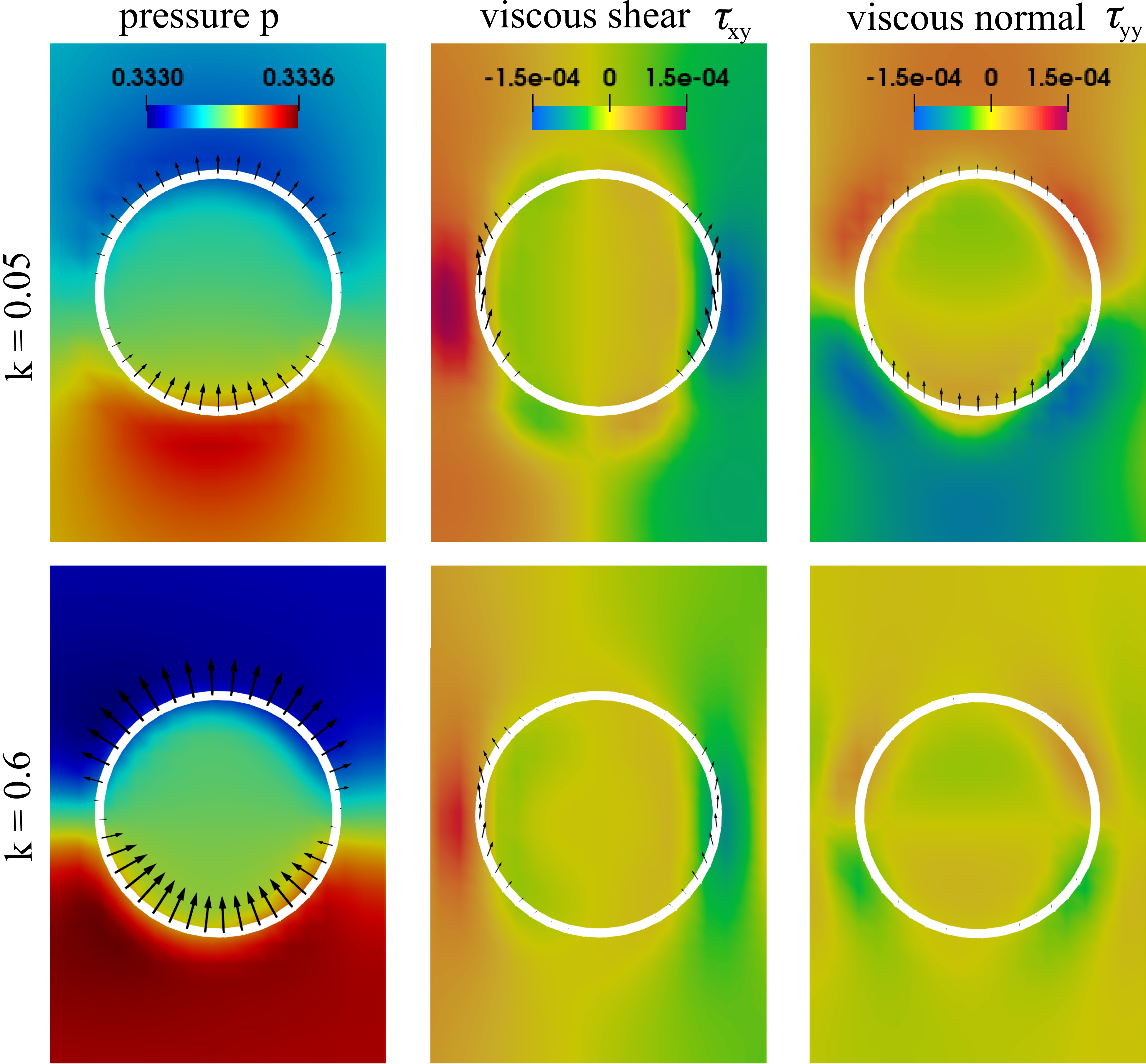}
  \caption{Pressure $p$ and viscous stress fields $\tau_{yx}$ and $\tau_{yy}$ for a capsule (white circle) along with the corresponding imparted forces on the membrane (black arrows). The capsule is at a $\mathrm{Bo}=0.079$. The top row is for a capsule at $k=0.05$ and bottom row for a capsule at $k=0.6$. The forces are represented as black arrows and their magnitude by the size of the arrows. All arrows are in the same scale.}
\label{fig:capsulepressureandstressfieldsconfinement}
\end{figure}

 We first consider the effect of confinement without flexibility. To do this we perform simulations for a stiff capsule at a $\mathrm{Bo}$ = 0.079. We consider the fields affecting drag. We illustrate the pressure $p$, normal $\tau_{yy}$ and shear $\tau_{yx}$ viscous stress components at $k=0.05$ and $k=0.6$ in Fig. \ref{fig:capsulepressureandstressfieldsconfinement}. Fig. \ref{fig:capsulepressureandstressfieldsconfinement} indicates that there is an increase in pressure gradient acting across the capsule boundary for a confined system. There is a stronger pressure gradient at $k=0.6$ than at $k=0.05$ due to the proximity of the walls. These add a retarding effect to the capsule. There is a noticeable decrease in the velocity gradient leading to lower values of $\tau_{xx}$ and $\tau_{yx}$. Since we know that $p$, $\tau_{yy}$ and $\tau_{yx}$ affect the drag, an increase or decrease in these components will change it. We plot the drag $\lambda_t$ as function of the confinement parameter in Fig. 3d of the main manuscript. 
We further measure the contributions of the viscous shear and viscous normal stresses to the drag force. We present in Table \ref{tab:dragcontributions} the values of the drag force and the respective contributions (along with pressure) at two different confinement levels. We see that for $k=0.05$, the largest contributor to drag is the viscous normal stresses, followed by viscous shear stresses and pressure. However, at $k=0.6$, the walls introduce strong pressure gradients making pressure the largest contributor to drag. 

\begin{table}
\caption{Individual contributions of pressure and viscous stresses to the drag force $\lambda_t$ (l.u).}
\begin{center}
    \begin{tabular}{ | c | c | c | c | c |}
      \hline
      \thead{Dim. size \\ $k$} &   \thead{Pressure \\ $p$} & \thead{Viscous \\ shear $\tau_{yx}$} & \thead{Viscous \\ normal $\tau_{yx}$} & \thead{Drag \\ $F^D$} \\
      \hline
      0.05 &  $3.7 \, 10^{-3}$  & $3.0\, 10^{-3}$  & $3.1 \, 10^{-3}$ & $9.8 \,10^{-3}$\\
      \hline
      0.6 &  $6.9 \, 10^{-3}$   & $2.1 \, 10^{-3}$  & $7.7 \, 10^{-4}$ & $9.8 \, 10^{-3}$\\
      \hline
    \end{tabular}
  \end{center}
\label{tab:dragcontributions}
\end{table}

\begin{figure}
  \includegraphics[width=1.0\linewidth]{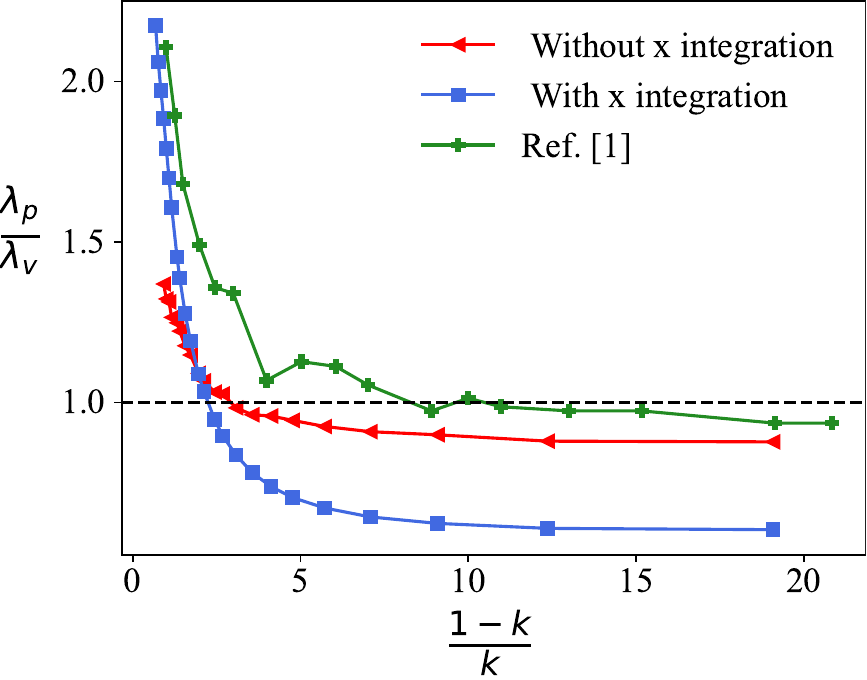}
  \caption{Comparison between results with (blue line) and without (red line) the $x$ coordinate integration. The green line corresponds to numerical values obtained in Ref. \cite{BenRichou_Ambari_Lebey_Naciri_2005} for further comparison. Dashed line to guide the eye}
  \label{fig:no_x_integration}
\end{figure}

Reference \cite{BenRichou_Ambari_Lebey_Naciri_2005} reports that for a solid cylinder (circular cross section) the ratio $\lambda_p / \lambda_v$ tends to 1 and tends to 1/2 for a sphere. We find values converging to $0.6$. The discrepancy is attributed to the numerical method. Because the capsule is  divided into nodes, small numerical errors can make it wiggle and not stay  exactly at $x=0$. We performed the same simulations without the $x$ integration of the capsule nodes i.e. nodes can only move in the $y$-direction. This fixes the $x$ component of the center of mass of the capsule. We find now that the ratio converges to $\approx 0.9$ (as seen in Fig. \ref{fig:no_x_integration}). An exact value of 1 is not achievable due to the interpolation nature of the method used to calculate the stresses and discritization of the capsule.
Lastly, we show that the capsule experiences negligible lift $F_x$ when compared to drag $F_y$ . Lift is caused by pressure $p$ and the stress components responsible for lift are $\tau_{xx}$ and $\tau_{xy}$. We reiterate that it is the $x$ component of the forces that contributes to lift. We note that these forces are symmetric and thus the net force is zero in the $x$-direction. Clearly due to interpolation this is three orders of magnitude smaller than the drag and therefore negligible.
\begin{figure}
  \includegraphics[width=0.55\linewidth]{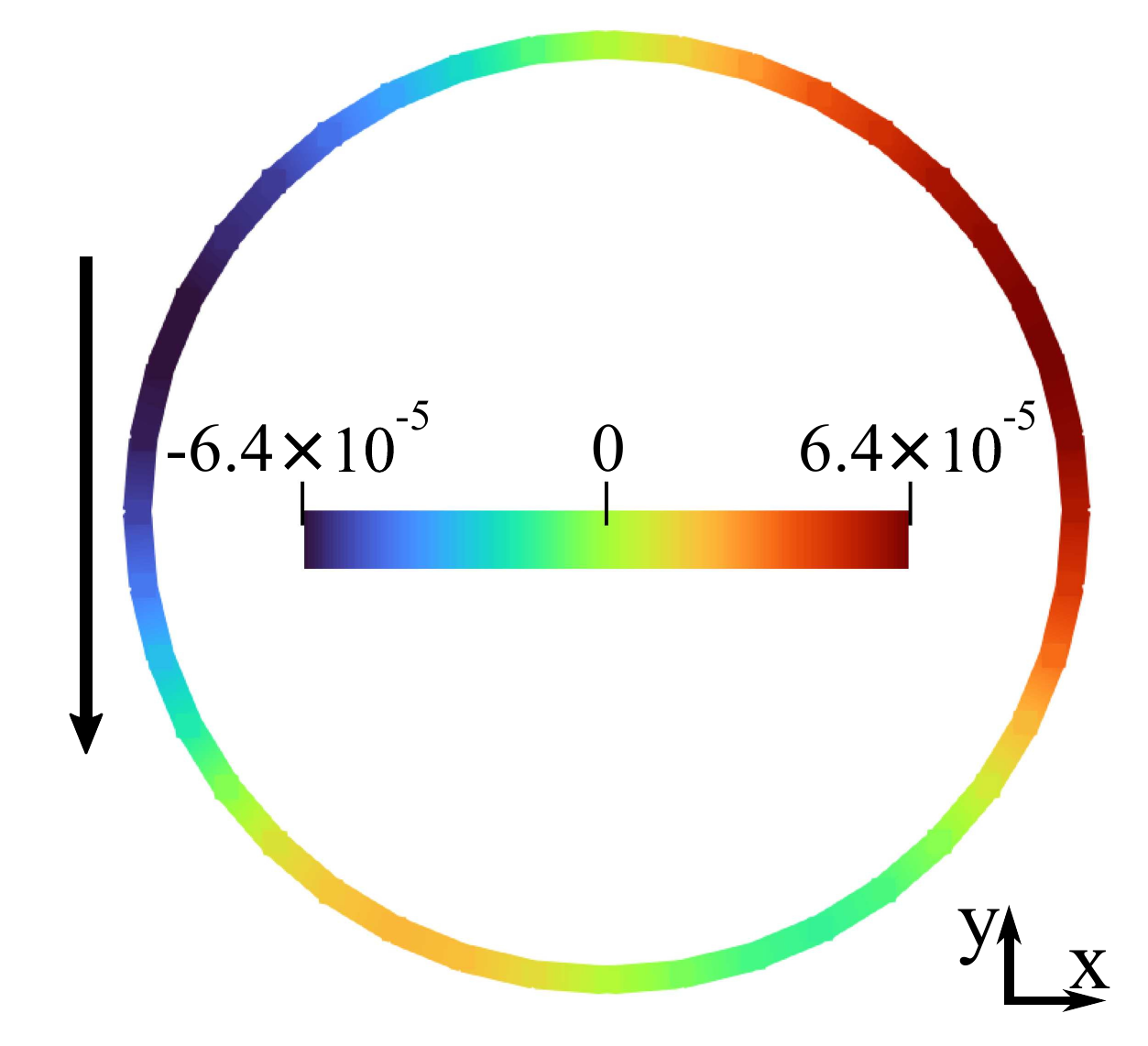}
  \caption{$x$ component $T_x$ of the traction vector. Contributions are $p$, $\tau_{xx}$ and $\tau_{xy}$. There is symmetry with respect to a vertical axis passing through the middle of the capsule. Since each half experiences symmetric forces the capsule does not move along the $x$-direction i.e. there is no lift. }
  \label{fig:liftsimmetry}
\end{figure}

\textcolor{black}{For completeness, we note that the monotonic increase of aspect ratio at small $k$ and its 
decrease at large $k$ saturate at finite values, set respectively 
by the balance of viscous stresses with elasticity ($k \to 0$) and by lubrication 
pressure in the narrow gaps ($k \to 1$).}

\subsection{Drag force: Viscous vs pressure contributions}
In Fig. \ref{fig:lambda}a, we show that as $k \to 1$ for a quasi rigid particle ($\mathrm{Bo} = 0.079$), the pressure contributions are significantly higher than their counterparts in Fig. \ref{fig:lambda}b.
After a crossing point, we notice that capsules with progressively higher $\mathrm{Bo}$ have increasing pressure contributions and decreasing viscous contributions. The pressure contours indicate that at high confinement (large $k$) the pressure difference inside and outside the capsule is greater than at low confinement (low $k$). Physically, the fluid particles are being compressed in the front-facing surface and extended in the back-faced surface.

\subsection{Effect of Capillary number}

\begin{figure}[H]
  \includegraphics[width=1.0\linewidth]{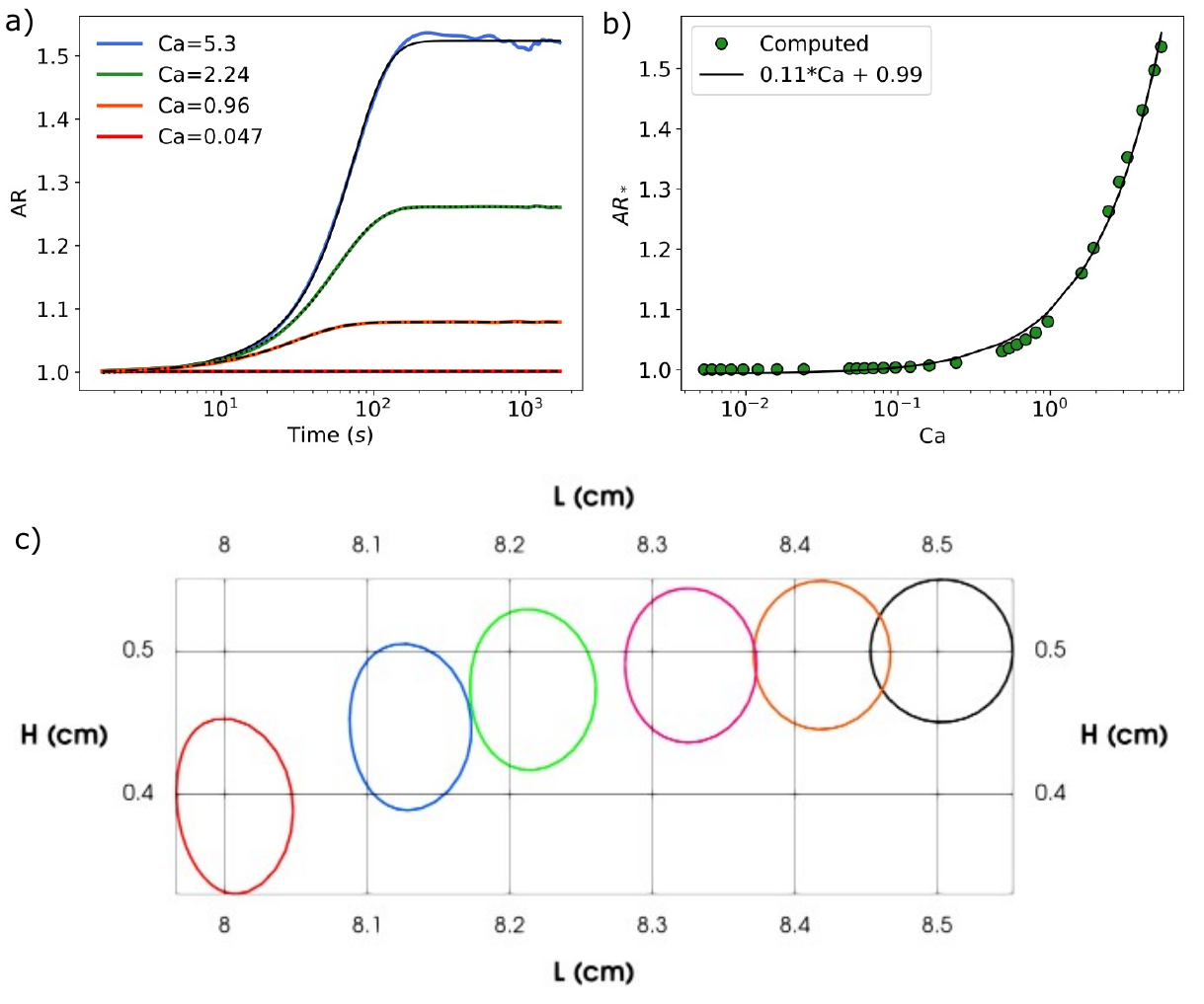}
  \caption{(a) Temporal evolution of AR for increasing values of Ca corresponding to more deformable membranes. (b) Asymptotic values of AR as a function of the aspect ratio with the corresponding curve fit. (c) Overlap of the final shape and position for different values of Ca. From the left to the right Ca = 5.2, Ca = 3.1, Ca = 2.34, Ca = 1.56, Ca = 0.93 and Ca = 0.005. }
  \label{fig:ca}
\end{figure}

\textcolor{black}{To explore the role of membrane deformability, we studied the dynamics of capsules for different values of the capillary number, $Ca = \mu V_t / k_s$ ,
which quantifies the competition between viscous stresses and membrane elasticity. Figure~\ref{fig:ca}(a) depicts the temporal evolution of the aspect ratio (AR) for increasing values of $Ca$. For small $Ca$, the AR remains close to unity, indicating nearly circular capsules. As $Ca$ increases, the AR increases until it reaches a stationary value $AR^\ast$, reflecting stronger elongation of the capsule. Figure~\ref{fig:ca} (b) and (c) compare the final steady-state shapes and positions for several values of $Ca$. Higher $Ca$ leads to larger deformations, with the capsules elongating perpendicular to their direction of motion. Additionally, capsules with higher $Ca$ are displaced less than their rigid counterparts, consistent with the increased dissipation associated with deformation. Together, these results demonstrate that the capillary number provides a key control parameter for capsule deformation and its dynamical response.  }

\subsection{Droplet simulations}
We also performed sedimenting droplet simulations using multicomponent lattice Boltzmann and the methodology reported in \cite{Silva_Coelho_daGama_Araujo_2023,coelho2023}. Here we vary the surface tension parameter as described in \cite{Silva_Coelho_daGama_Araujo_2023}. We found that the shape changes are very similar. Due to the nature of the pseudopotential, it is not possible to vary too strongly the surface tension without nonphysical effects manifesting such as spurious velocities. In addition, we attempted to measure the pressure and viscous stresses acting across the diffusive interface. To identify the interface, we calculate the normal vector $\hat{n} = \frac{\nabla \rho}{|\nabla \rho|}$. However, because of the interface is diffuse, it is difficult to calculate the viscous stresses acting on the droplets. For droplets, we used the same methodology reported in Ref. \cite{Silva_Coelho_daGama_Araujo_2023} A droplet with 12 l.u diameter (compared to 12.5 for capsules) is placed in a fully bounded 2D domain with parameters $G_{AB}=G_{BA}=3.0$, $G_{k,1}=-7.9$, $G_{k,2} = 4.9$ with a density of 1.2 lu. These droplet is also placed at $(x_0, y_0) = (0.5W, 0.89L)$. We applied an external force to the bulk of the droplet to drive sedimentation. The results are depicted in Fig. 1 of the main text, which is in line with the experiments on droplets.
\begin{figure}[H]
  \includegraphics[width=1.0\linewidth]{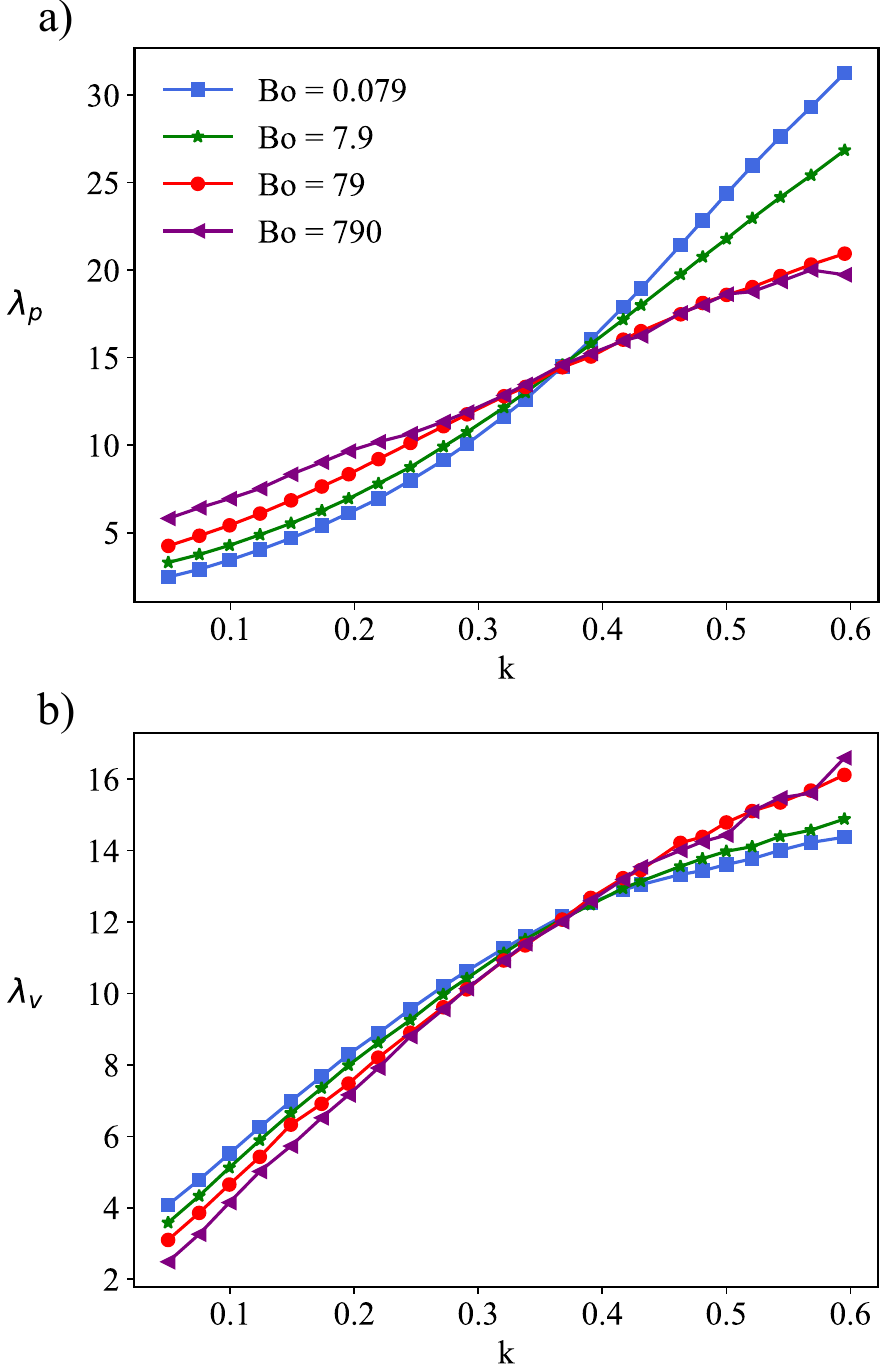}
  \caption{(a) Pressure contributions and (b) viscous contributions to drag.}
  \label{fig:lambda}
\end{figure}

\section{Rigid cylinder near a wall in the lubrication limit}

A cylinder of radius $a = D/2$ is moving with velocity $U_0$ a distance $h_0$ from a wall,
where $h_0 = (W - D)/2$. The distance between the cylinder and the wall is denoted $h(x)$.
We assume $h_0/a \ll 1$. The cylinder creates a flow $(u,v)$. In this lubrication limit,
we assume that the horizontal flow $u$ is much more substantial than the vertical one $v$,
i.e., $u \gg v$. We can also approximate the shape of the cylinder in this limit as a parabola
since far terms will not contribute. In dimensionless distance, $H(x) = h(x)/h_0$, the distance
to the wall is~\cite{Leal2007}:
\begin{equation}
H = 1 + \frac{a}{h_0} - \frac{a}{h_0} \sqrt{1 - \left( \frac{x}{a} \right)^2}
\approx 1 + X^2,
\end{equation}
where we have defined a dimensionless horizontal distance $X = x/l$, and
$l = \sqrt{2 a h_0}$. The parameter $l$ is a typical horizontal length in the problem.
We introduce dimensionless variables:
\begin{align*}
    U &= \frac{u}{U_0}, & Y &= \frac{y}{h_0}, & X &= \frac{x}{l},
    & P &= \frac{p}{p_c}, & \text{with } p_c = \frac{\mu U_0 l}{h_0^2}.
\end{align*}
The boundary conditions in these variables are:
\begin{equation}
U(Y = 0) = -1, \quad U(Y = H) = 0.
\end{equation}
Assuming $\text{Re} \ll 1$, $h_0/l = \epsilon \ll 1$ and $U \gg V$, the Navier--Stokes equations
in the lubrication approximation take the form:
\begin{align}
    & 0 = - \frac{\partial P}{\partial X} + \frac{\partial^2 U}{\partial Y^2}, \label{eq:NS1} \\
    & P = P(X), \label{eq:NS2} \\
    & \frac{\partial U}{\partial X} + \frac{\partial V}{\partial Y} = 0. \label{eq:NS3}
\end{align}
Solving for $U$ from Eq.~\eqref{eq:NS1} using Eq.~\eqref{eq:NS2}:
\begin{equation}
U = \frac{1}{2} \frac{dP}{dX} Y^2 + A Y + B.
\end{equation}
Using the boundary conditions, we find $B = -1$ and
$A = \frac{1}{H} - \frac{1}{2} \frac{dP}{dX} H$. Thus, the velocity profile is:
\begin{equation}
U = \frac{1}{2} \frac{dP}{dX} Y (Y - H) + \frac{Y}{H} - 1.
\end{equation}
The pressure gradient is determined from the incompressibility condition:
\begin{equation}
dV \vert _0 ^H = -\int_0^{H(X)} \frac{\partial U}{\partial X} dY.
\end{equation}
Since $V(H) = V(0) = 0$, applying Leibniz's rule, we obtain:
\begin{equation}
0 = \frac{\partial}{\partial X} \int_0^{H(X)} U dY.
\end{equation}
Using the velocity equation in the integral above, we derive an equation for the pressure:
\begin{equation}
C = - \frac{1}{12} \frac{dP}{dX} H^3 - \frac{H}{2}.
\end{equation}
Approximating the height as $H(X) = 1 + X^2$, we integrate for the pressure:
\begin{equation}
P = P_0 + \int \frac{A - 6(1 + X^2/2)}{(1 + X^2/2)^3} dX.
\end{equation}
With $P(\infty) = P(-\infty) = P_0$, we obtain $A = 8$, leading to:
\begin{equation}
P = P_0 + \frac{8X}{(X^2 + 2)^2}.
\end{equation}
This profile is antisymmetric around the particle’s center: there is high pressure in front
of the particle and low pressure behind it. When the particle is no longer rigid, these pressures
deform it into the bullet-like shape observed in experiments and simulations. The shear stress
is given by $\mu \partial u / \partial y$. Comparing it with the typical pressure
$P_c = \mu U_0 l / h_0^2$, and using $\epsilon = h_0/l$, we obtain:
\begin{equation}
\frac{P_c}{\mu U_0 / h_0} = \frac{1}{\epsilon} \gg 1.
\end{equation}
Thus, in the lubrication limit, pressure dominates over shear stress.

{\color{black}\section{Origin of the critical confinement $k \approx 0.37$}

The observed crossover at $k \approx 0.37$ marks the change from a viscous-shear-dominated
hydrodynamic regime to a pressure-dominated regime. Physically, when the particle is sufficiently
close to the confining walls, the thin gap produces large pressure gradients in front of and behind
the particle (lubrication pressure) that grow faster with confinement than the viscous-shear stresses.
This change in the dominant stress explains why the steady shape switches from oblate-like to
bullet-like at a well-defined $k$.

A compact scaling argument based on the lubrication limit (see the previous section) gives a similar value  for the critical confinement. Let the cylinder radius be $a = D/2$ and the
half-gap to the wall be $h_0 = (W-D)/2$. In the lubrication approximation, the characteristic
lubrication length is $l \sim \sqrt{2 a h_0}$ and the characteristic pressure scales as
\[
P_c \sim \frac{\mu U\, l}{h_0^2} \sim \mu U \frac{\sqrt{2a h_0}}{h_0^2}
= \mu U \sqrt{\frac{2a}{h_0^3}} \,,
\]
while a characteristic viscous shear scales as
\[
\tau_{\mathrm{visc}} \sim \frac{\mu U}{h_0}.
\]
Taking the ratio using $\varepsilon = h_0/l = \sqrt{h_0/(2a)}$ yields
\[
\frac{P_c}{\tau_{\mathrm{visc}}} \sim \frac{1}{\varepsilon} = \sqrt{\frac{2a}{h_0}}.
\]

Pressure therefore becomes comparable to (and then exceeds) viscous shear when
$\sqrt{2a/h_0} \sim 1$, i.e., when $h_0 \sim 2a$. Expressing $h_0$ and $a$ in terms of the
confinement $k = D/W$,
\[
h_0 = \frac{W-D}{2} = \frac{W(1-k)}{2}, \qquad
a = \frac{D}{2} = \frac{kW}{2},
\]
the condition $h_0 \sim 2a$ reduces to
\[
\frac{1-k}{2k} \sim 1,
\]
or $k \sim 1/3$. This simple estimate therefore predicts a crossover $k_c$ of order $0.33$, in
good agreement with the numerically observed value $k \approx 0.37$. The small difference is
expected because the scaling neglects finite-size corrections and
numerical interpolation/smoothing inherent to the IBM.

The crossover confinement $k_c \approx 0.37$ identified in the simulations corresponds to the
balance of pressure and viscous stresses for capsules in two dimensions. This threshold is not
expected to be observed in the present droplet experiments, as the interfacial tension maintains
nearly spherical or bullet-like shapes across the accessible range of $k$. The analytical scaling
estimate included here confirms the robustness of this value in the simulated regime.

\subsection{Dependence on other parameters and confidence}

Within our study we kept the density and viscosity ratios equal to unity (experiments and
simulations) and operated at low Reynolds numbers; under these conditions the crossover location
is robust and essentially independent of flexibility (Bond number), as demonstrated in Fig.~2
(the shape transition line is invariant across Bo) and confirmed by the crossing of pressure and
viscous drag in Fig.~3. The scaling shows why this is so: viscosity $\mu$ and characteristic
velocity $U$ largely cancel in the nondimensional comparison of pressure vs.~viscous stress, so
moderate changes in those parameters do not strongly affect the geometric condition that sets $k_c$.

Nevertheless, we emphasize two caveats. (i) Large changes in viscosity or density ratios
(for example, a strongly viscous inner fluid compared to the outer fluid) or substantially higher
Reynolds numbers could alter the relative magnitudes of the stress contributions and shift $k_c$.
(ii) The lubrication estimate is derived for a near-wall cylinder in the small-gap limit;
three-dimensional spherical geometry, finite-domain effects, and the diffuse representation of
interfaces in the droplet simulations and IBM may produce modest quantitative shifts. For these
reasons we interpret $k \approx 0.37$ as a robust threshold for the conditions studied here (low Re,
unit viscosity/density ratio), consistent between Figs.~2 and 3. Mapping the sensitivity of $k_c$
to large variations of density/viscosity ratio or to higher Re is a natural direction for future work.

}


\bibliography{apssamp}
\bibliographystyle{aipauth4-1}
\end{document}